\documentclass[sigconf,screen]{acmart}
\usepackage{multirow}
\usepackage{verbatim}
\usepackage{geometry}
\usepackage{subfig}
\usepackage{enumitem}
\usepackage{colortbl}
\usepackage{xcolor}
\usepackage{diagbox}

\newcommand{\getcolor}[3]{%
    \pgfmathsetmacro{\colorValue}{100.0*(#1-#2)/(#3-#2)}%
    \edef\temp{\noexpand\cellcolor{blue!\colorValue}\textcolor{black}{#1}}%
    \temp
}
\AtBeginDocument{%
  }

\setcopyright{acmlicensed}
\copyrightyear{2025}
\acmYear{2025}
\setcopyright{cc}
\setcctype{by}
\acmConference[CHI '25]{CHI Conference on Human Factors in Computing
Systems}{April 26-May 1, 2025}{Yokohama, Japan}
\acmBooktitle{CHI Conference on Human Factors in Computing Systems (CHI
'25), April 26-May 1, 2025, Yokohama,
Japan}\acmDOI{10.1145/3706598.3713707.}
\acmISBN{979-8-4007-1394-1/25/04}

\begin{document}


\title[AAG and OPG]{Actual Achieved Gain and Optimal Perceived Gain: Modeling Human Take-over Decisions Towards Automated Vehicles' Suggestions}

\author{Shuning Zhang}
\orcid{0000-0002-4145-117X}
\authornotemark[1]
\email{zsn23@mails.tsinghua.edu.cn}
\affiliation{%
  \institution{Tsinghua University}
  \city{Beijing}
  \country{China}
}

\author{Xin Yi}
\orcid{0000-0001-8041-7962}
\authornote{These authors contributed equally to this work.}
\email{yixin@tsinghua.edu.cn}
\affiliation{
    \institution{Tsinghua University}
    \city{Beijing}
    \country{China}
}
\affiliation{
    \institution{Zhongguancun Laboratory}
    \city{Beijing}
    \country{China}
}

\author{Shixuan Li}
\orcid{0009-0008-6828-6347}
\email{li-sx24@mails.tsinghua.edu.cn}
\affiliation{
    \institution{Tsinghua University}
    \city{Beijing}
    \country{China}
}

\author{Chuye Hong}
\orcid{0009-0005-4679-2212}
\email{hongcy21@mails.tsinghua.edu.cn}
\affiliation{%
  \institution{Tsinghua University}
  \city{Beijing}
  \country{China}
}

\author{Gujun Chen}
\orcid{0000-0001-5071-1175}
\email{cgj24@mails.tsinghua.edu.cn}
\affiliation{%
  \institution{Tsinghua University}
  \city{Beijing}
  \country{China}
}
\author{Jiarui Liu}
\orcid{0009-0007-8937-7128}
\email{15774169531@163.com}
\affiliation{%
  \institution{Tsinghua University}
  \city{Beijing}
  \country{China}
}
\author{Xueyang Wang}
\orcid{0000-0002-9797-9491}

\email{wang-xy22@mails.tsinghua.edu.cn}
\affiliation{
    \institution{Tsinghua University}
    \city{Beijing}
    \country{China}
}
\affiliation{
    \institution{Zhongguancun Laboratory}
    \city{Beijing}
    \country{China}
}

\author{Yongquan Hu}
\orcid{0000-0003-1315-8969}
\email{yongquan.hu@unsw.edu.au}
\affiliation{
    \institution{University of New South Wales}
    \city{Sydney}
    \country{Australia}
}

\author{Yuntao Wang}
\orcid{0000-0002-4249-8893}
\authornote{Corresponding author.}
\email{yuntaowang@tsinghua.edu.cn}
\affiliation{
    \institution{Tsinghua University}
    \city{Beijing}
    \country{China}
}

\author{Hewu Li}
\orcid{0000-0002-6331-6542}
\email{lihewu@cernet.edu.cn}
\affiliation{
    \institution{Tsinghua University}
    \city{Beijing}
    \country{China}
}
\affiliation{
    \institution{Zhongguancun Laboratory}
    \city{Beijing}
    \country{China}
}

\renewcommand{\shortauthors}{Zhang et al.}
\begin{abstract}
    
    Driver decision quality in take-overs is critical for effective human-Autonomous Driving System (ADS) collaboration. However, current research lacks detailed analysis of its variations. This paper introduces two metrics--Actual Achieved Gain (AAG) and Optimal Perceived Gain (OPG)--to assess decision quality, with OPG representing optimal decisions and AAG reflecting actual outcomes. Both are calculated as weighted averages of perceived gains and losses, influenced by ADS accuracy. Study 1 (N=315) used a 21-point Thurstone scale to measure perceived gains and losses—key components of AAG and OPG—across typical tasks: route selection, overtaking, and collision avoidance. Studies 2 (N=54) and 3 (N=54) modeled decision quality under varying ADS accuracy and decision time. Results show with sufficient time (>3.5s), AAG converges towards OPG, indicating rational decision-making, while limited time leads to intuitive and deterministic choices. Study 3 also linked AAG-OPG deviations to irrational behaviors. An intervention study (N=8) and a pilot (N=4) employing voice alarms and multi-modal alarms based on these deviations demonstrated AAG's potential to improve decision quality.

\end{abstract}





\begin{teaserfigure}
\centering \includegraphics[width=0.9\textwidth]{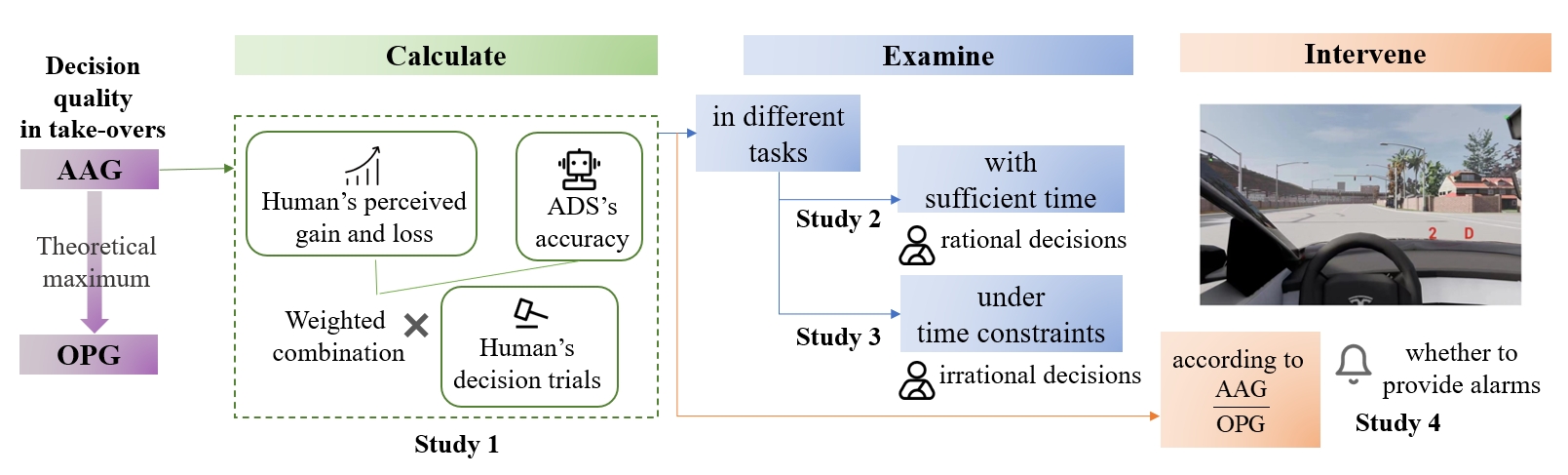}
    \caption{The research framework of this paper.}
    \label{fig:aag_OPG}
\end{teaserfigure}

\maketitle
\section{Introduction}

Although automated vehicles (AVs) have made significant progress and can handle various driving scenarios \cite{hancock2020challenges}, they still face challenges in managing all road situations. As a result, human drivers must occasionally take control, especially in complex or unexpected scenarios where AV systems reach their limits \cite{du2020predicting}. This ongoing need for human intervention highlights its essential role in human-vehicle collaboration during AVs' transition towards full autonomy.

Most research on automated driving focuses on full transfer of control when systems reach their operational limits \cite{zhang2019determinants,bazilinskyy2018take}. However, the take-over process is often suboptimal due to human factors, such as reduced situational awareness \cite{vlakveld2018situation}. Walch et al. \cite{walch2017car} proposed cooperative interfaces to address these challenges, emphasizing mutual predictability, control, shared awareness, and calibrated trust. While studies have examined trust calibration \cite{akash2020toward,kraus2020more}, situational awareness \cite{woide2022effect,walch2019cooperative}, and control~\cite{colley2021investigating}, mutual predictability, particularly the ability of AVs to predict driver behavior, remains under-explored. 



Previous studies on mutual predictability have mainly examined the quality of drivers' take-over under various conditions, using metrics like acceleration \cite{lee2021effects}, deviation distance \cite{pakdamanian2021deeptake}, and subjective reports \cite{du2020predicting} after the take-over. However, these studies overlook the crucial role of decision-making before the actual control. These decisions are crucial for responding safely and effectively to ADS suggestions, especially in scenarios where the driver only partially takes over and relies on ADS support or road analysis \cite{walch2016towards}. In such cases, decision quality becomes a key factor, as it directly impacts the outcome. Therefore, \textbf{this paper aims to quantitatively model the quality of drivers' decision-making during take-over in response to ADS suggestions.}

We conceptualized human and ADS decision-making as a collaborative system, aiming for optimal collaborative outcomes. As control shifts to human drivers, their thoughts and decisions directly influence the take-over quality. To predict real-time take-over quality without relying on post-hoc ground-truth, we propose evaluating outcomes based on drivers' perceptions of benefits and risks, offering a new perspective. As decision-making quality may be influenced by factors like time~\cite{wan2018effects}, take type~\cite{radlmayr2014traffic} and ADS accuracy~\cite{wan2018effects}, we selected these factors for their representativeness of human decision-making, ADS performance, and environmental conditions. We examined the effects of these key factors on take-over quality to validate our proposed metrics. This leads to the following research questions: 

\noindent \textbf{RQ1.} How to establish quantitative metrics grounded in the human-ADS collaboration perspective to assess drivers' decision-making quality during the take-over process?

\noindent \textbf{RQ2.} How do time, task and ADS accuracy factors influence drivers' take-over decision quality?

\noindent \textbf{RQ3.} How to improve drivers' take-over decision quality leveraging the proposed quantitative metrics?

To address RQ1, we developed a pair of correlated metrics—Actual Achieved Gain (AAG) and Optimal Perceived Gain (OPG)—to assess the quality of human decisions in response to ADS suggestions. OPG, grounded in game theory and the concepts of perceived gain and loss, represents the optimal decisions drivers make to maximize perceived gains across various scenarios. Perceived gains and losses serve as indicators of decision quality, reflecting participants' value judgments. It is calculated as the maximum weighted sum of perceived gains and losses for each scenario, with weights determined as the ADS accuracy and the sum encompassing all possible decision behaviors. The evaluation of perceived gain and loss utilizes a 21-point Thurstone scale based on behavioral economics principles \cite{mohring1971alternative,kahneman2013prospect}, and reflects participant preferences. However, due to time constraints, drivers often struggle to make fully rational decisions. Therefore, we proposed AAG to quantify the actual value drivers achieve through their decisions, capturing potential deviations from OPG.

In Study 1, we measured perceived gain and loss across three typical take-over tasks with varying safety risks: Route Selection~\cite{sheng2021trust}, Overtaking~\cite{gold2015trust,borojeni2016assisting,lin2020understanding,zeeb2015determines} and Avoiding Collision with Other Cars~\cite{louw2017you,pampel2019getting,melcher2015take}. These parameters are integral to AAG and OPG, linking human cognition with decision-making and ADS suggestions. Our findings reveal that perceived gain and loss vary by task, with safety-preserving decisions (e.g., avoiding collisions when suggested so) showing significantly higher perceived gain and loss than others (e.g., taking a shorter route when suggested so), indicating a tendency towards risk aversion and follow-the-suggestion behavior~\cite{rabin2013risk}.

To address RQ2, Studies 2 and 3 investigated the effectiveness of AAG and OPG under conditions of sufficient and insufficient decision-making time. Study 2 manipulated \textit{ADS accuracy} and \textit{task factors}. Using the CARLA simulator~\cite{dosovitskiy2017carla}, we constructed automated driving scenarios with varying levels of ADS accuracy and decision-making time to assess drivers' decisions responses to ADS suggestions. The results indicated that given sufficient time, drivers' AAG converged towards OPG, reflecting drivers made optimal decisions aimed at maximizing gains. Study 3, on the other hand, varied \textit{time} and \textit{task factors} to explore scenarios where decision-making time was insufficient. Interestingly, under these time constraints, drivers exhibited a tendency to make simple, risk-averse decisions (e.g., consistently avoiding collisions) regardless of ADS recommendations, resulting in a divergence between AAG and OPG. We attributed this deviation to risk aversion and a tendency towards conservative decision-making. The results showed the potential leveraging AAG \& OPG to improve decision quality. 

To address RQ3, Study 4 preliminarily evaluated AAG's effectiveness by developing a prototype using voice alerts~\cite{wen2024adaptivevoice}. The alerts were designed with three levels of urgency based on the degree of AAG's deviation from OPG. As the deviation increased, the alerts were more urgent, effectively capturing drivers' attention. With acceptable sacrifice in cognitive load, these differentiated alerts improved AAG relative to OPG, enhancing decision-making. In a follow-up experiment, we explored a multimodal approach combining voice alerts with visual cues. While the results were preliminary, the findings suggest that multimodal cues may further support driver decision-making, offering valuable insights for future research.

In summary, our contributions are three-folded:

\begin{itemize}
    \item We introduced the first quantitative metrics-Actual Achieved Gain (AAG) and Optimal Perceived Gain (OPG), grounded in game theory, to model and assess drivers' decision quality during take-overs.
    \item We applied AAG and OPG to three representative take-over tasks, examining the effects of task type, time, and ADS accuracy on these metrics. Our findings highlight that limited time leads to more irrational or conservative decisions.
    \item We demonstrated the practical application of AAG through interventions using voice alarms and multimodal cues, which were adjusted according to the AAG/OPG deviation.
\end{itemize}

\section{Background \& Related Work}
This work focused on the take-over process during semi-automated driving. We first presented the overview of methods and factors modeling the take-over process. We then introduced the works focused on perceived gain and loss in autonomous driving. Finally we presented works around interaction design during take-over.

\subsection{Modeling the Outcome of Take-over Process}

Recent studies have highlighted the criticality of take-over in AVs, where human drivers must resume control~\cite{pakdamanian2021deeptake}. During take-over, previous research indicates that drivers often face challenges in preparing for manual control, underscoring the importance of understanding their take-over performance~\cite{hubmann2017decision,zeeb2015determines}. Factors influencing this performance include the difficulty of non-driving related tasks (NDRT), drivers' characteristics, and road conditions~\cite{du2020predicting,pakdamanian2021deeptake}. Notably, previous models primarily focused on aspects of lateral control, neglecting the role of decision-making accuracy in take-over performance. Understanding how drivers make decisions in these moments is crucial, particularly as advanced ADS are being developed to assist drivers' take-over~\cite{boelhouwer2019should}. For example, Boelhouwer et al.~\cite{boelhouwer2019should} found providing a system manual during take-over did not help drivers. Leitner et al.~\cite{leitner2023overtake} explored how time pressure and the criticality of the situation affect drivers' overtaking decisions, but only under road repairing task. In comparison, we examined drivers' behavior in different take-over tasks and proposed the metric to reflect human's take-over quality without ground truth results. 

The factors influencing take-over performance encompass both objective (e.g., tasks, ADS accuracy, decision time)~\cite{zeeb2015determines} and subjective factors (e.g., drivers' personality, environmental conditions)~\cite{korber2016influence,ma2021drivers,radlmayr2014traffic,wandtner2018effects}. This study specifically examines objective factors, acknowledging that subjective factors predominantly influence driving processes through objective means. However, previous works such as the work by Zeeb et al.~\cite{zeeb2015determines} only investigated the effect of NDRT on take-over time, without examining drivers' decision. 

Objective factors such as task complexity significantly impact take-over performance, often mediated by risk perception~\cite{scharfe2020impact, gold2016taking, radlmayr2014traffic}. Scharfe et al.~\cite{scharfe2020impact} investigated the effect of drivers' familiarity of the task and the traffic's objective complexity on take-over quality, reflected only through subjective ratings. Gold et al.~\cite{gold2016taking} investigated the effect of traffic density on take-over time, which was reflected through take-over time and lateral, longitudinal capabilities. Radlmayr et al.~\cite{radlmayr2014traffic} examined the traffic situations and NDRT on drivers' take-over time. However, they did not take the inherent risk and behavior into the calculation of the take-over quality. This study selects three tasks with varying risk perceptions to explore their corresponding impacts. Additionally, the reliability of ADS is crucial, affecting drivers' perception of system reliance and their responsiveness to system-initiated take-over requests~\cite{KORBER201818,schwarz2019effect}. Decision accuracy and time are also critical; insufficient decision time may lead to rushed decisions, compromising their quality~\cite{walch2017autonomous,radlmayr2015haptic,gold2013take,zhou2021does}. We also examined the effect of decision time on the proposed metrics, which reflect drivers' decision quality.

\subsection{Perceived Gain and Loss in Automated Driving}

Perceived gain and loss, grounded in behavioral economics~\cite{mohring1971alternative,kahneman2013prospect}, captures how humans assess potential benefits and risks in decision-making contexts. Similar dynamics of benefit and risk have been observed in various daily tasks, such as time savings~\cite{lin2015customer} and life safety~\cite{oliver2018your}, however did not adopt the framework in autonomous driving scenarios. Kool et al.~\cite{kool2017cost} applied the gain and loss framework to human decision-making in a two-step task, but their research did not address collaborative decision-making with AI systems. Vasconcelos et al.~\cite{vasconcelos2023explanations} applied the gain and loss framework to examine human trust and overreliance on AI, particularly in maze-solving tasks. Unlike these studies, which emphasize cognitive demands, task difficulty, and monetary incentives as gains and losses, our work focus on safety and time considerations in autonomous driving. We aim to model human decision quality within this context, exploring factors such as time pressure, which is particularly relevant to autonomous driving. 

In human-ADS collaboration, several studies have adopted the perceived gain and loss perspective. The past literature mainly used perceived gain and loss to analyze the acceptance of automated driving. For example, Brell et al.~\cite{brell2019scary} used questionnaires to assess drivers' risk perception, perceived benefits and barriers in autonomous driving. They found risk perceptions for conventional driving were significantly smaller compared with automated driving. Mertens et al.~\cite{mertens2020need} and Bearth et al.~\cite{bearth2016risk} also derived similar results, contributing to the marketing of AVs. These past studies primarily focused on the acceptance perspective rather than detailed decision-making processes.

Li et al.~\cite{li2024investigating} found the positive correlation between perceived benefit and ADS's capability and the mediating effect of perceived benefit on trust. However, they lacked the quantitative analysis between perceived gain and drivers' decisions. Additionally, they adopted a qualitative perspective in understanding the perceived gain and loss of automated driving. In contrast, we propose AAG and OPG for modeling and explaining drivers' take-over decisions.

\subsection{Interaction Design of Take-over Process}

The interaction design during take-over included the interface~\cite{yun2020multimodal}, layout and the presented information~\cite{cohen2017effects}. Modalities were categorized into voice~\cite{koo2016understanding}, visual~\cite{faltaous2018design} and tactile~\cite{manawadu2016haptic}. Although visual feedback is vivid, it often disrupts other non-driving related tasks (NDRTs)~\cite{prewett2011meta}. Tactile feedback requires specialized seats and is not widely implemented on the vehicles. Voice feedback is preferred as it imposes low mental load, allowing drivers to complete NDRTs simultaneously~\cite{iqbal2011hang}.

Based on the aforementioned studies, we used voice feedback for suggestions during take-over requests. During take-over, the ADS presented hazard situations~\cite{cohen2017effects, yang2018hmi} (i.e. the task) and the suggestions~\cite{telpaz2015haptic, eriksson2018rolling, langlois2016augmented, lindemann2019exploring, lorenz2014designing} to drivers. While during the beginning of the collaborative driving, the ADS introduced its accuracy~\cite{helldin2013presenting, kunze2019automation, white2019rebuilding}, the urgency level (i.e., the decision time)~\cite{politis2015language, roche2018should} to drivers. 

Given the urgent nature of take-over events, presenting complex information can overwhelm drivers. While earlier approaches focused on alarms, recent advancements integrate various information cues to facilitate smoother transitions to manual control~\cite{forster2017driver,eriksson2018rolling,kim2023and}. These cues include detection accuracy, hazard warnings, suggested lane changes, and intended directions or actions~\cite{helldin2013presenting,kunze2019automation,white2019rebuilding,cohen2017effects,yang2018hmi,telpaz2015haptic}. Some studies further emphasize the urgency to prompt quicker driver responses~\cite{politis2015language, roche2018should}. This study aligns with these trends by integrating suggestions, accuracy, and decision urgency to enhance human-ADS interaction and situation awareness~\cite{beller2013improving,manger2023providing}.

\section{Actual Achieved Gain and Optimal Perceived Gain}

During take-over, ADS typically provided suggestions to human drivers~\cite{boelhouwer2019should,telpaz2015haptic,eriksson2018rolling,lindemann2019exploring}. Drivers make driving decisions prioritizing safety and time efficiency. Drawing from behavioral economics~\cite{kahneman2013prospect}, we proposed to use ``perceived gain and loss'' to reflect drivers' preference for safe and time-saving choice. Leveraging these parameters, we further introduced Actual Achieved Gain (AAG) and Optimal Perceived Gain (OPG) to enable the modeling of drivers' decision rationality and quality, thereby improving ADS's understanding of drivers. 

\subsection{Definition and Calculation of AAG and OPG}\label{sec:def}

The basis of AAG and OPG is perceived gain and loss. \textbf{Perceived gain and loss is defined as ``driver's perception of what they could benefit from a specific driving decision (e.g., braking when the ADS ask them to brake) regarding the safety and time saving perspective'' similar to thew past literature \cite{brell2019scary,li2019no}. } It is measured through subjective scales following the measurement of risk and time saving perception~\cite{brell2019scary}. Zero on the scale indicates no perceived gain and loss, corresponding to the case where no take-over decisions happened and ADS operate normally. The maximum and minimum value represented extreme situations such as causing or saving a life, or significant time savings or losses. 

\textbf{AAG} quantifies decision quality by summing the weighted average of perceived gains and losses, with weights determined by ADS accuracy (see Formula~\ref{equ:aag}). Here, $p$ represents ADS accuracy, and $PG_{n, D, V}$ denotes the perceived gain or loss associated with specific driver and ADS decisions ($D$ and $V$) over $n$ decision instances. $\bar{PG}$ reflects gains and losses when the ADS suggests an alternative action to the driver's choice, incorporating ADS inaccuracies (Table~\ref{tbl:perceived_gain}). For example, when the ADS suggests avoidance and the driver complies, $PG_{D, V}$ applies; if the driver instead avoids despite an ADS suggestion not to, $\bar{PG_{D,V}}$ is used. A higher AAG indicates a stronger alignment with optimal decisions (Figure~\ref{fig:aag_OPG}). This metric, initially defined for single tasks, can be generalized to multiple tasks by summing the respective AAG values.
\textbf{OPG} represents the theoretical maximum of AAG, calculated as the weighted sum of perceived gains and losses, with weights determined by ADS accuracy (see Formula~\ref{equ:OPG}). It reflects the optimal decision-making scenario. However, drivers may deviate from this ideal due to irrational behaviors (see Section~\ref{sec:study3_OPG}). The deviation between AAG and OPG indicates reduced decision quality, with both metrics sharing similar influencing factors and formulas, except that OPG represents the optimal situation.

\begin{equation}\label{equ:aag}
    AAG = \sum_{n = 1}^{N} (p PG_{n, D, V} + (1-p)\bar{PG}_{n, D, V})
\end{equation}

\begin{equation}\label{equ:OPG}
    OPG = \max_{D, V} \sum_{n=1}^{N}(p PG_{n, D, V}+(1-p)\bar{PG}_{n, D, V})
\end{equation}

\subsection{Measurement and Usage of AAG and OPG}

AAG and OPG are grounded in the principle that decision quality can be quantified through perceived gains and losses ($PG$), reflecting how drivers evaluate risks and benefits in varying scenarios. Through evaluating the perceived gain and loss, these metrics could help capture the interplay between human decision-making and ADS performance.

OPG represents the optimal decision-making quality, assuming rationality and full situational awareness, while AAG measures actual decisions made under real-world constraints. The discrepancy between the two reveals deviations in decision quality, which can stem from time pressure, cognitive biases, or task complexity. These results may vary according to tasks, time and ADS accuracy, which were separately examined.

AAG and OPG are designed to improve human-ADS collaboration. For example, when significant deviations between AAG and OPG are detected, targeted interventions, such as adaptive alarms, can guide drivers toward safer and more optimal decisions. By linking decision-making patterns to situational awareness and ADS performance, AAG and OPG provide actionable insights to enhance both system design and user interaction strategies.

\section{Study 1: Measuring Perceived Gain and Loss for AAG \& OPG}

We conducted a large-scale questionnaire study (N=315) to assess perceived gains and losses for calculating AAG in various take-over scenarios, while also examining drivers' cognition of take-over decisions to gain deeper insights into their decision-making quality.

\subsection{Participants and Apparatus}

We recruited 315 participants (160 males, 155 females, age range from 18 to 65, M=32.2, SD=6.8) in China through distributing the questionnaire on an online platform \footnote{\url{https://www.wjx.cn}}. The demographics of the participants was shown in Table~\ref{tbl:demo}. The average driving experience was 7.1 years (SD=5.6). We ensured the diversity in driving experience, driving frequency, vehicle types, residential provinces and ages, etc. 

\begin{table*}[htbp]
    \centering
    \begin{tabular}{p{3cm}|p{12cm}}
    \toprule
        \textbf{Age} & 2 within 15-20, 40 within 20-25, 71 within 25-30, 126 within 30-35, 41 within 35-40, 18 within 40-45, 12 within 45-50, 2 within 50-55, 2 within 55-60, 1 within 60-65 \\ \midrule
        \textbf{Education} & 75 with high school degrees or below, 181 with bachelor degrees, 45 with master degrees, 14 with Ph.D. degrees \\ \midrule
        \textbf{Driving experience} & 40 have driving experience of over 10 years, 117 have that of 5--10 years, 75 have that of 1--3 years, 83 have that of equal or less than 1 year \\ \midrule
        \textbf{Driving frequency} & 129 drive many times each day, 119 drive one time each day, 57 drive 1 time each week and 10 drive less than 1 time each week \\ \midrule
        \textbf{Vehicle types} & 313 drive cars with Manual Transmission (MT), 18 drive cars with Auto Transmission (AT), 2 with trucks / buses \\
    \bottomrule
    \end{tabular}
    \caption{Participants' demographics for Study 1. The sum of the vehicle types is greater than 315 because one participant could drive multiple types of vehicles.}
    \Description{}
    \label{tbl:demo}
\end{table*}
To ensure data quality, we applied three control methods for participants: 1) Participants had to spend more than $4N$ seconds on the questionnaire, where $N$ is the number of questions. 2) Participants were required to pass random test questions inserted into the questionnaire. 3) Participants’ answers had to align with common sense, such as the perceived loss from a collision being greater than that from not colliding. These checks excluded 31 out of 346 participants (8.95\%), resulting in 315 valid participants. Each participant received 30 RMB for completing the questionnaire. We randomly selected 30 participants (10\%) to conduct an exit interview after the experiment.

\subsection{Questionnaire Design}\label{sec:study1_design}

\textbf{We measured the perceived gain and loss of three typical take-over tasks by describing the cases to participants and having them fill in scales.} The task refers to the specific action or decision drivers must take during take-over. Take-over tasks are mainly categorized into: obstacles~\cite{gold2015trust} (e.g., stranded cars~\cite{gold2015trust}, road construction), travel planning changes~\cite{xin2019literature,kuehn2017takeover} and capability limits~\cite{xin2019literature,louw2017you} (e.g., autonomous system's failure~\cite{xin2019literature}, potential collision~\cite{louw2017you}). We selected three common tasks considering different levels of risk: Avoid Collision with Other Vehicles (denoted as ``Avoid Collision'')~\cite{kim2017takeover,wan2018effects,fitch2011driver} with the highest risk, Overtake~\cite{wan2018effects} with medium risk and Route Selection~\cite{tan2022computational} with low risk. Following previous literature~\cite{boelhouwer2019should}, each task was designed as a binary choice. \textbf{Participants were presented with the following descriptions for each task, including potential outcomes and judgment criteria:}

\textbf{Avoid Collision with Other Vehicles}: We informed participants that \textit{``In this task, you need to avoid the oncoming traffic at the intersection ahead. Assuming you have just reached the intersection, the ADS provides suggestions based on whether a vehicle is detected in your blind spot of your lateral field of vision. ADS could help you make the judgment. Avoiding oncoming traffic may avoid traffic accidents, while if there is no actual collision, not avoiding will save time. ''} ADS would suggest ``Avoid'' or ``Not avoid'', while participants could take one choice.

\textbf{Overtake}: We informed participants that \textit{``In this task, the ADS is driving on a regular road with slow-moving vehicles ahead. The ADS suggests either choosing the left lane to overtake the front car or not overtake it. ADS suggestions will mainly be based on the driving behaviour of other potential cars behind on the left side, which was connected through the Internet of Things (IoT). In this situation, ADS could potentially help you during decision. If you choose to overtake when overtaking is not suggested by the ADS, there may be traffic accidents during the overtaking process.''} ADS would suggest ``Overtake'' or ``Not overtake'', while participants could take one choice.

\textbf{Route Selection}: We informed participants that \textit{``In this task, ADS is driving on an unfamiliar road and there are two options for the upcoming route. The ADS suggest either the actually longer or shorter route based on its judgment. The actual shorter route may be slower due to traffic congestion or road repairs, while the actual longer route may be faster due to a smoother traffic flow. You need to make the decision between these two options by considering ADS's suggestions.''} ADS would suggest ``Select long route'' or ``Select short route'', while participants could take one choice.

We carefully selected the experimental materials to ensure clarity in scenario comprehension and improve the reliability of questionnaire responses. A preliminary study compared three presentation formats: video-based, text-only, and image-text combinations, with 15 participants per condition. Feedback showed that participants understood and envisioned potential threats in all formats. The text-only condition encouraged deeper reflection, while the video format risked drawing attention to specific details and was perceived as less realistic. Based on score consistency, we chose the text-based approach for its clarity and comprehensiveness, which better supported participants' understanding.

Participants were encouraged to consider potential outcomes beyond the provided scenarios. However, they indicated that the outcomes provided reflected what they would typically consider in real-life or work situations. Their assessments were based on the ADS recommendations and hypothetical decisions instead of the visual stimuli to avoid random effects.

We used Thurstone scales (ranging from -10 to 10)~\cite{thurstone1927method} for measurements, with descriptions to participants as in Section~\ref{sec:def}. We provided participants with decision options (avoid, not avoid, etc.) and the ADS suggestions (avoid, not avoid, etc.) for each rating item. Participants evaluated the perceived gains and losses towards each situation we presented (2 ADS suggestions $\times$ 2 human decisions $\times$ 3 tasks $=$ 12 entries). The ADS was assumed to be 100\% accurate in its predictions. Detailed questionnaire information is provided in Section~\ref{sec:questionnaire} of the appendix. 


To better understand drivers' following and risk aversion~\cite{holt2002risk}, in addition to the gain and loss values, we examined participants' perceptions of risk, time savings, and trust in ADS. We introduced two metrics: Following Gain of Task and Choice Gain of Task, detailed in Section~\ref{sec:study1_cognition}.

\subsection{Procedure}

We first informed each participant of the study and the setting of the questionnaire. We then let participants sign the consent. They then fill all entries of perceived gain and loss with different tasks and situations. The entire session lasted 12 minutes. We randomly selected 30 participants (about 10\%) from the total 315 to interview about their reasons behind the rating, especially in terms of perceived gain and loss. 

\subsection{Results}
We first examined whether the data is valid through calculating inter-rater reliability. The overall Cronbach's alpha is 0.766 and for each question the Cronbach's alpha is over 0.7, indicating a strong consistency (0.7-0.9). We then conducted statistical testing towards the data, where we first examined whether the data conformed to normal distribution. For these data which conformed to normal distribution, we used Repeated Measures Analysis of Variance (RM-ANOVA) test for statistical analysis. Otherwise, we used Friedman non-parametric test. 

\subsubsection{Perceived Gain and Loss}\label{sec:study1_perceived}

Table~\ref{study1_task} presented the average perceived gain and loss across tasks and situations. Perceived gains were highest in safety-sensitive scenarios, such as avoiding collisions, while perceived losses were greatest when avoidance failed. In the "Route Selection" task, some participants showed no clear preference, resulting in similar perceived gains and losses for both shorter and longer routes. Notably, tasks with lower traffic accident risks corresponded to lower perceived gains from accident avoidance.  Consequently, the perceived gain for the "Short Route" was lower than for "Avoid Collision." Time savings also significantly influenced ratings, with ratings for "Long Route," "Overtake," and "Not Avoid" decreasing as time-saving benefits diminished.

\begin{table}[ht]
    \centering
    \caption{Perceived gain and loss of different conditions with different tasks. The number in parenthesis denoted one standard deviation. D denotes the drivers' decision, while A denotes the ADS suggestion.}
    \label{tbl:perceived_gain}
    
    \subfloat[Route selection.]{%
        \resizebox{0.25\textwidth}{!}{
        \begin{tabular}{c|c|c}
            \toprule
            \diagbox{D}{A} & short route & long route \\
            \midrule
            short route & 3.59 (1.85) & -0.22 (1.95) \\
            long route & -1.92 (2.00) & 4.15 (1.78) \\
            \bottomrule
        \end{tabular}%
        }
        \label{tbl:route}
    }
    \subfloat[Overtake.]{%
    \resizebox{0.25\textwidth}{!}{
        \begin{tabular}{c|c|c}
            \toprule
            \diagbox{D}{A} & overtake & not overtake \\
            \midrule
            overtake & 3.92 (1.91) & 0.55 (1.89) \\
            not overtake & -2.74 (2.17) & 3.72 (1.87) \\
            \bottomrule
        \end{tabular}%
        }
        \label{tbl:overtake}
    }
    \vspace{1ex}  
    \subfloat[Avoid collision.]{%
    \resizebox{0.25\textwidth}{!}{
        \begin{tabular}{c|c|c}
            \toprule
            \diagbox{D}{A} & avoid & not avoid \\
            \midrule
            avoid & 5.57 (1.68) & 0.25 (2.06) \\
            not avoid & -3.96 (2.06) & 2.77 (2.14) \\
            \bottomrule
        \end{tabular}%
        }
        \label{tbl:avoid}
        \hspace{0.3cm}
    }
    \subfloat[Annotation.]{%
    \resizebox{0.21\textwidth}{!}{
        \begin{tabular}{c|c|c}
            \toprule
            \diagbox{D}{A} & choice 1 & choice 2 \\
            \midrule
            choice 1 & $PG_{00}$ & $PG_{01}$ \\
            choice 2 & $PG_{10}$ & $PG_{11}$ \\
            \bottomrule
        \end{tabular}%
        }
        \label{tbl:annotation}
    }
    
    \Description{This table shows the perceived gain and loss of different conditions with different tasks (a) Route Selection, (b) Overtake, (c) Avoid Collision, (d) The annotation. The number in parenthesis denoted one standard deviation.}
    \label{study1_task}
\end{table}

\subsubsection{Human's Cognition of Take-over Situations}\label{sec:study1_cognition}

We examined perceived gain and loss across various take-over situations and decisions to better capture human decision-making quality, analyzing from two perspectives: \textbf{Following Gain of Task} and \textbf{Choice Gain of Task}.

\textbf{Following Gain of Task} refers to the difference in perceived gain between following and not following the ADS, calculated as $PG_{11}+PG_{00}-PG_{10}-PG_{01}$ (see Table~\ref{study1_task}). This metric reflects participants' \textit{relative perceived gain} from following ADS suggestions. Notably, \textit{relative perceived gain may exceed 10 due to comparisons between opposite choices}. The highest value was for `Avoid Collision' (12.06), followed by `Overtake' (9.83) and `Route Selection' (9.87). An RM-ANOVA showed significant task differences ($F_{2, 628} = 13.7$, $p < .001$), highlighting a stronger perceived benefit in `Avoid Collision' when following ADS suggestions.

\textbf{Choice Gain of Task} is defined as $PG_{10}+PG_{11}-PG_{01}-PG_{00}$, reflecting participants' \textit{relative perceived gain} when making a specific decision. For `Avoid Collision', this metric was the highest (7.01), significantly greater than for `Overtake' (3.49) and `Route Selection' (1.14), as indicated by a Friedman non-parametric test ($\chi_2^2=27.0$, $p < .001$). This suggests a greater emphasis on safety, with less concern for route length. These indicators reflected participants have the preference for safer and time-saving choices, which paved the foundation of using AAG and OPG as the quality indicator.

\section{Study 2: AAG \& OPG with different ADS's Accuracy}

After collecting the perceived gain and loss, which is the core parameters of AAG and OPG, we analyzed how AAG and OPG reflected human's decision with varying ADS accuracy.

\subsection{Participants and Apparatus}
We recruited 54 Chinese participants (30 males, 24 females) using snowball sampling~\cite{goodman1961snowball}. Participants' age ranged from 18 to 54 and had a mean age of 28.3 years (SD=9.0) and an average driving experience of 5.8 years (SD=5.3). Participants were from varied occupations with varied educational backgrounds. Each participant received 100 RMB as compensation.

\begin{figure}[htbp]
    \subfloat[]{
    \includegraphics[width=0.9\columnwidth]{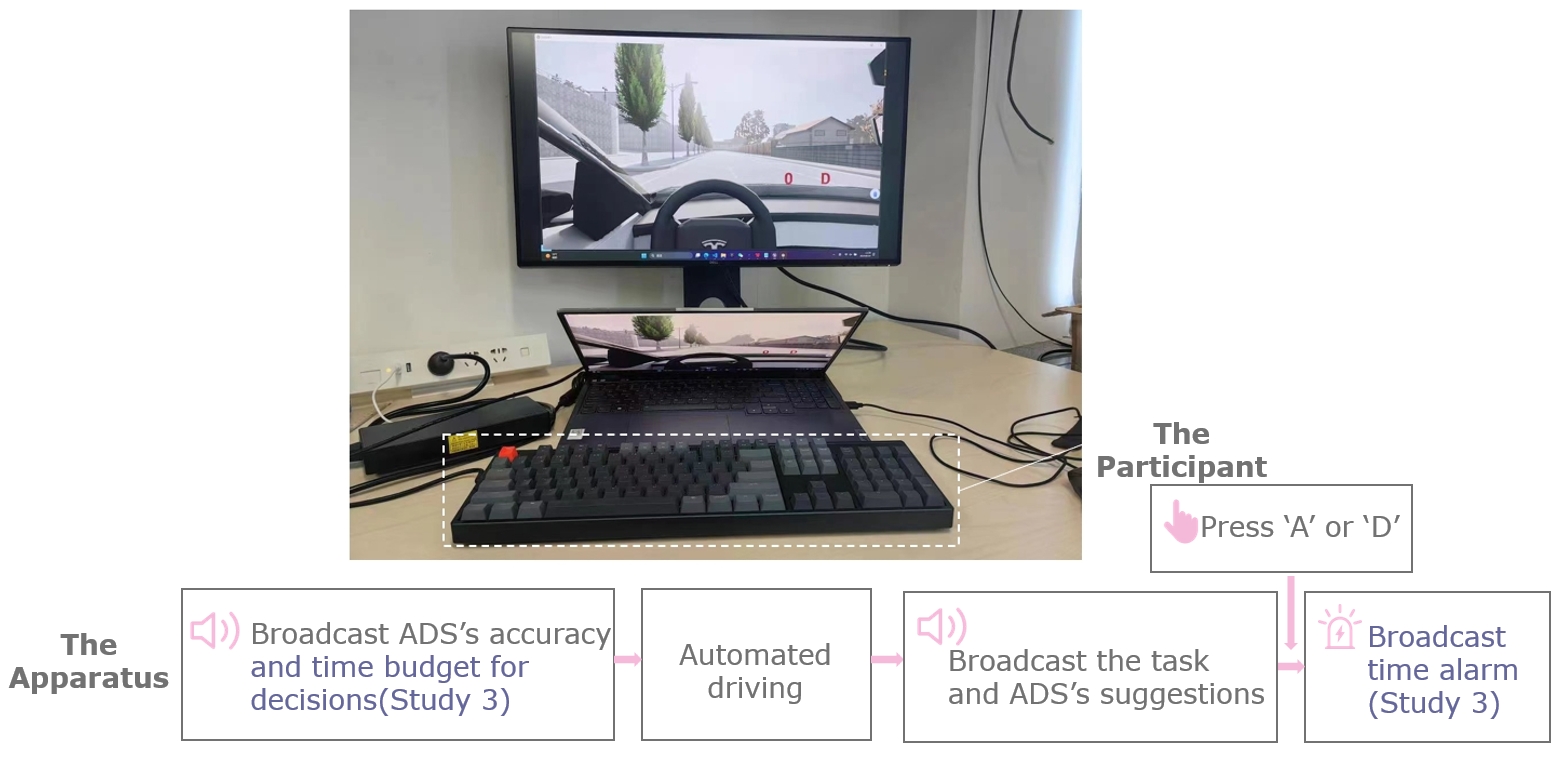}
    }
    \caption{The experiment platform for Study 2 and 3. The participants took a first-perspective view. (1) The system told participants of ADS's accuracy before the start of automated driving using voice. (3) The system told participants the suggestions of ADS when driving near the take-over intersection using voice. (4) The alarm indicating the time ran out in Study 3. When driving near the take-over intersection, the alarm rang. The ``0 D'' on the screen illustrated 0 miles per hour while ``D'' indicated gear level (forward gear). }
    \label{fig:platform}
    \Description{}
\end{figure}

For ethical and safety reasons~\cite{mcgehee2010perception}, we conducted the experiment in a simulator rather than a real vehicle, following previous studies~\cite{manawadu2015analysis}. The experiment was run on a 27-inch monitor with a 144 Hz refresh rate, chosen to minimize discomfort associated with virtual reality simulators~\cite{ihemedu2017simulation}. Participants used keyboard keys for decision-making instead of steering wheels and pedals, enabling faster, easier decisions. Figure~\ref{fig:platform} illustrates the experimental setup. We used the CARLA simulator~\cite{dosovitskiy2017carla} with its Python API for scene creation and interaction. Voice alerts were generated using Psytsx with default parameters~\cite{koo2016understanding,detjen2021increase}. The experiment was controlled using PyQt5 and system-level programming to ensure consistency across interaction rounds. The display refresh rate of 144 Hz ensured participant comfort throughout the experiment.

\subsection{Study Design}

\begin{figure}[htbp]
    \subfloat[Route Selection.]{
        \includegraphics[width=0.33\linewidth]{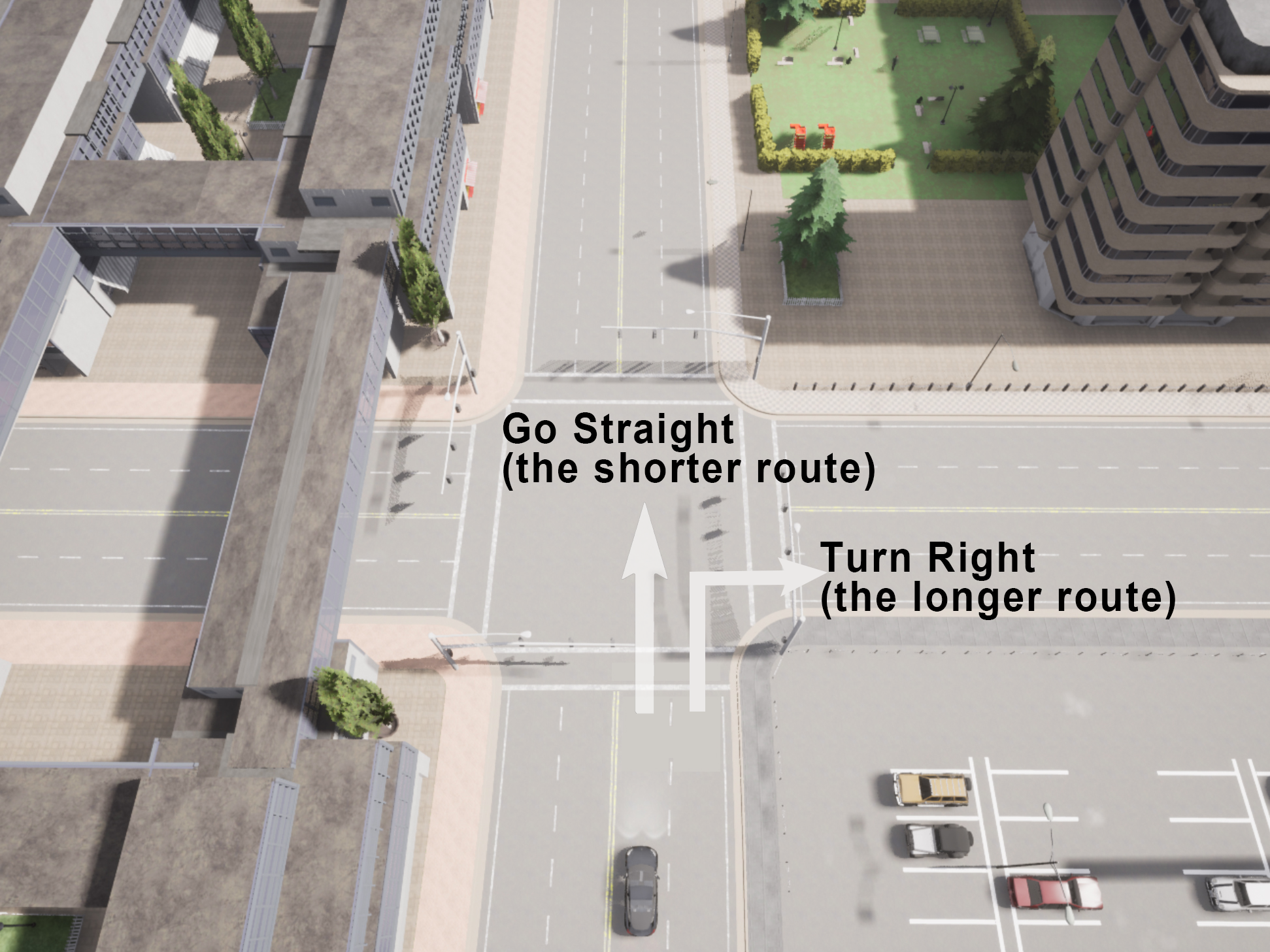}
        \label{fig:task_route_select}
    }
    \subfloat[Overtake.]{
        \includegraphics[width=0.33\linewidth]{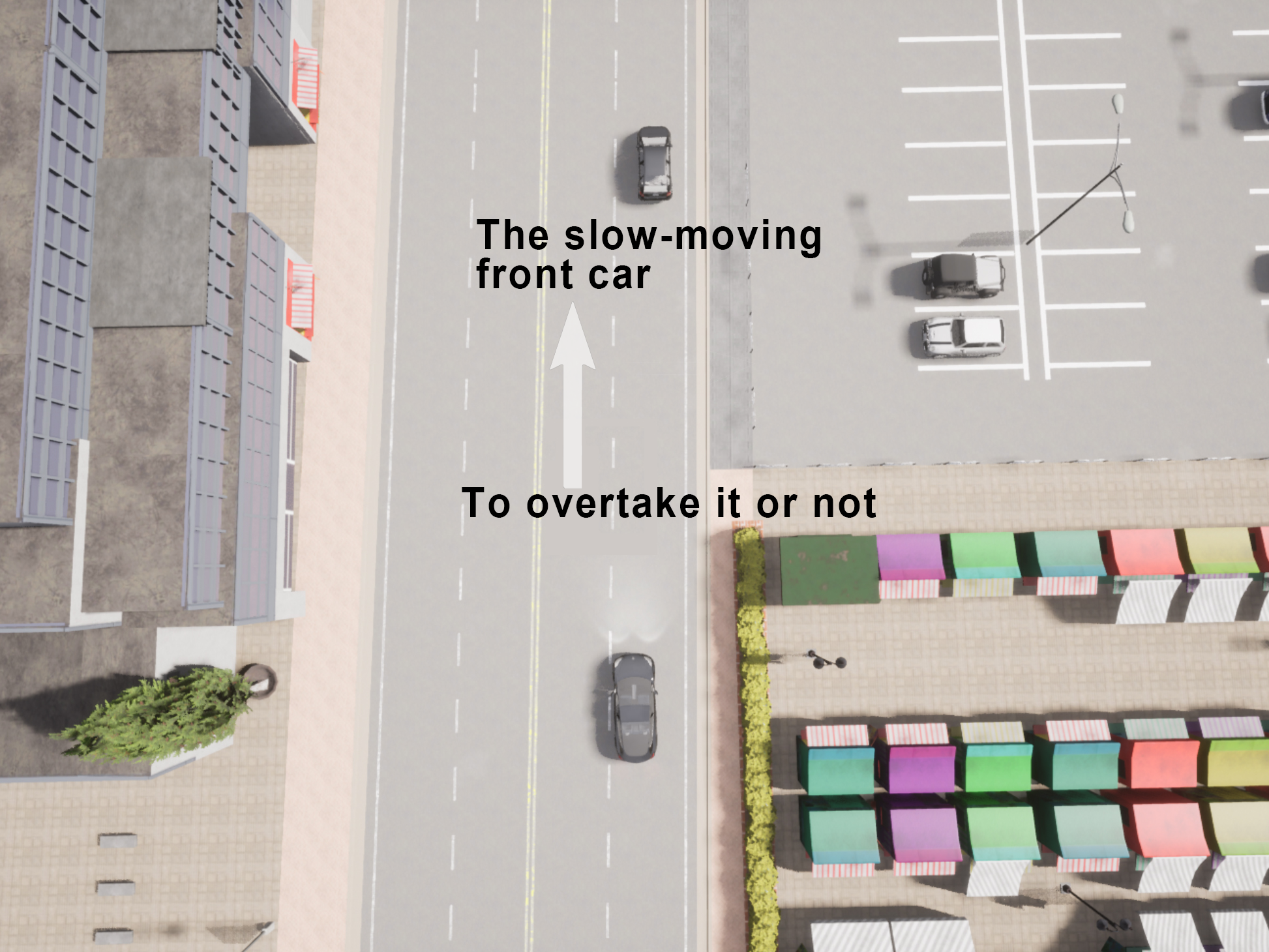}
        \label{fig:task_overtake}
    }
    \subfloat[Avoid Collision.]{
        \includegraphics[width=0.33\linewidth]{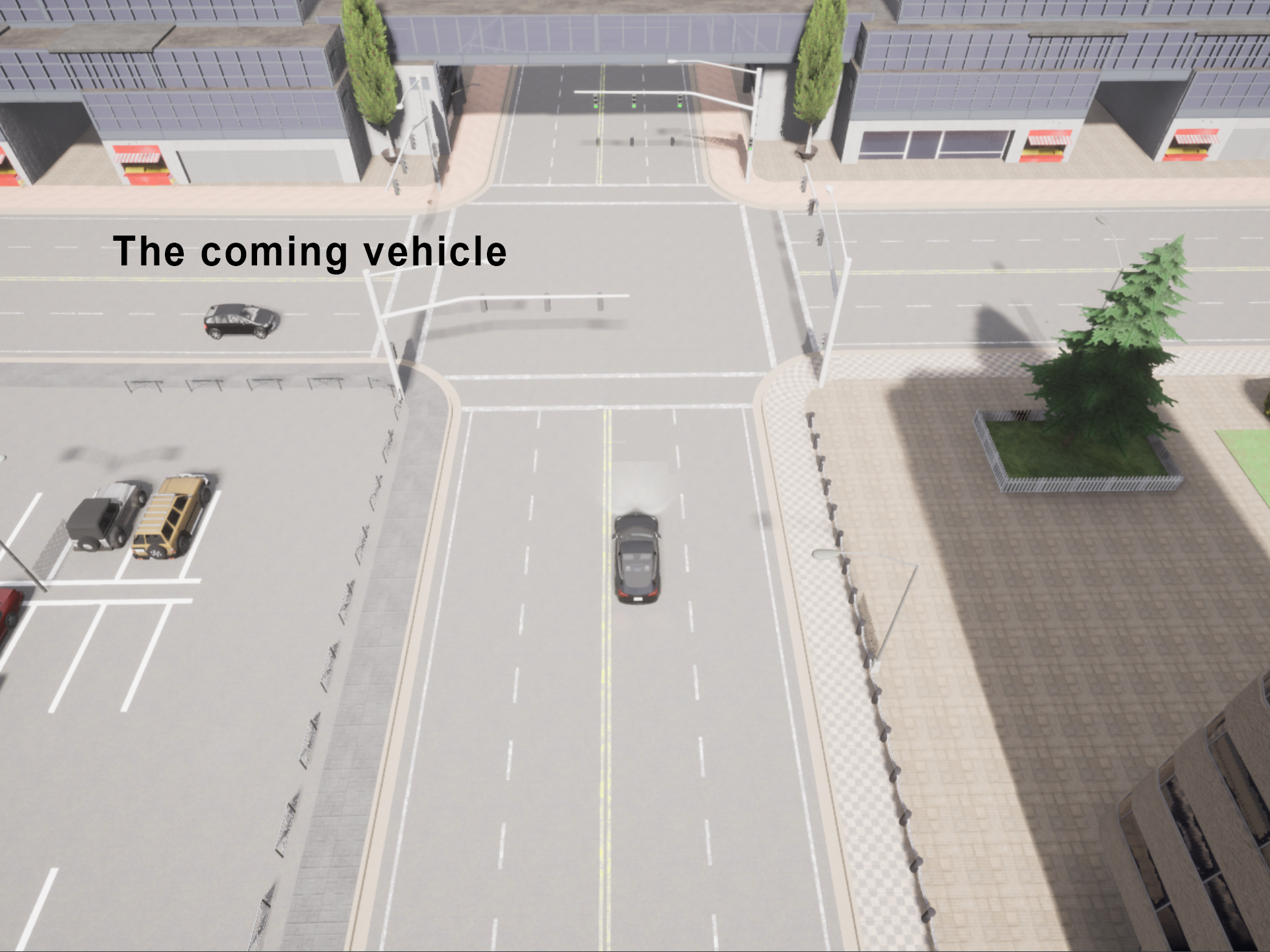}
        \label{fig:task_avoid}
    }
    \caption{Illustration of the different take-over tasks on the experiment platform. The arrows and text were added manually for clarification and were not part of the original system.}
    \Description{}
    \label{fig:illustration_task}
\end{figure}

We employed a two-factor within-subjects design, with \textit{task} and \textit{ADS accuracy} as factors, to assess the effectiveness of AAG and OPG across various take-over tasks: Route Selection, Overtake, and Avoid Collision (See Figure~\ref{fig:illustration_task}). 

ADS accuracy was defined as the percentage of correct judgments made by the ADS, without distinguishing between false alarms and misses, as drivers typically perceive overall accuracy. We set three levels of ADS accuracy: 60\%, 90\%, and 99\%, based on a pilot study with 24 participants, where participants considered 90\% as the minimum acceptable accuracy for reliable ADS operation, and 99\% as a reliable standard. Some companies also adopted this criteria\footnote{\url{https://www.sohu.com/a/491760010\_121136454}, accessed by Sep 12th, 2024}.

Participants collaborated with the ADS in four rounds per accuracy level and task, resulting in 36 trials. Each trial began from a standstill, with the ADS initiating autonomous driving, providing suggestions, and requesting a decision. Decisions were recorded by participants pressing `A' for ``Not Avoid'', ``Overtake'', or ``Select Short Route'', and `D' for ``Avoid'', ``Not Overtake'' and ``Select Long Route''. Decision time was measured from the end of the voice suggestion to the key-press timestamp. The ground truth (i.e., whether avoid means collision) was pre-set based on ADS's accuracy and was not disclosed to participants.

\subsection{Experiment Platform Setting}

The display showed real-time driving from a first-person perspective. Each trial began with the ADS verbally informing the participant of its recognition accuracy (``The recognition accuracy of the system is XX\%''). The ADS then initiated autonomous driving at a maximum speed of 30 miles per hour, driving to a randomly preset position for 15 to 60 seconds before describing the task and providing suggestions through voice. Participants were given sufficient time to develop rational strategies and made decisions by pressing the appropriate keyboard key after hearing the suggestions.

To mimic real-world scenarios and prevent habituation, participants were not informed of the task until just before they needed to take over. The ADS provided task details and suggestions simultaneously. Trials varied in length, lighting, road layout, scene, and maps in the simulator to avoid any learning effect and keep participants focused on the task rather than the environment.

To minimize the influence of previous trials on subsequent decisions, results of the decisions (e.g., whether a collision occurred) were not shown. After the ADS made suggestions, the display would freeze, proceeding to the next trial only after participants made their decisions.


\subsection{Procedure}

We briefed participants on the simulation platform and experiment process to ensure understanding, followed by a 3-minute familiarization period. The experiment consisted of 36 trials (2 suggestions × 2 repetitions × 3 tasks × 3 ADS's accuracy levels) split into two sessions, with trial order randomized using a Latin Square design~\cite{mckay2005number}. Each 40-second trial began with the ADS informing participants of its accuracy, followed by autonomous driving and decision-making. Decision-making time was measured from the initiation of take-over request to the key press, averaging 3.50 seconds (SD=0.89). A 90-second break was provided between sessions to reduce fatigue. Subjective feedback was collected after the experiment.

\subsection{Results}
We primarily analyze the AAG and OPG of drivers based on the perceived gain and loss. AAG was calculated based on the perceived gain and loss derived from Study 1, using the mean value of each participant's rating. Adopting the Equation~\ref{equ:aag} and substituting the drivers' decisions, the AAG reflected participants' preference. OPG was calculated by optimizing AAG for the specific task. We further analyzed the correlation between AAG and OPG in this study. Finally, we identified patterns in participants' decision-making behavior influenced by tasks and ADS accuracy.

\subsubsection{AAG Converge Towards OPG with Enough Decision Time}\label{sec:study2_OPG}

\begin{figure}[!htbp]
    \subfloat[]{
        \includegraphics[width=0.47\textwidth]{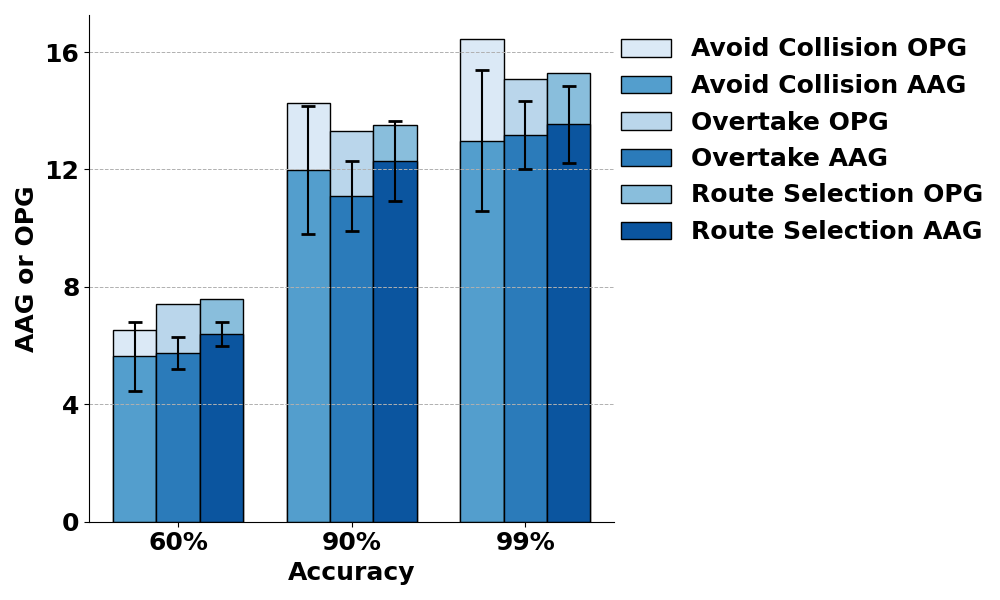}
        \label{fig:study2_aag}
    }
    
    \subfloat[]{
        \includegraphics[width=0.47\textwidth]{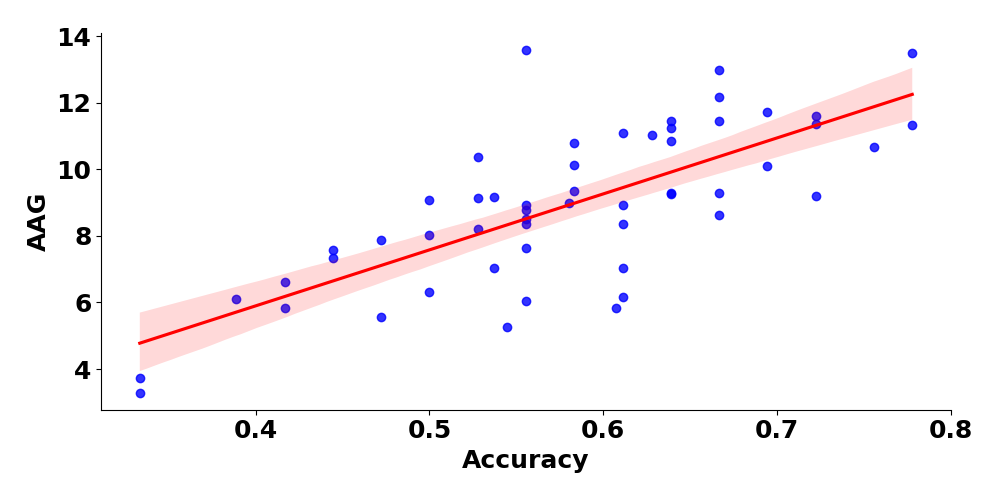}
        \label{fig:study2_correlation}
    }
    \caption{(a) AAG and OPG in different tasks with different accuracy. Errorbar indicated one standard deviation. The lighter bar indicated OPG while the darker bar indicated AAG. (b) The correlation between AAG and ground truth accuracy, where red share indicates 95\% confidence interval.}
    \Description{}
    \label{fig:study2_aag_opg}
\end{figure}

Figure~\ref{fig:study2_aag} displayed the AAG and OPG across different ADS accuracy levels and tasks. On average, AAG was only 15.4\% (SD=4.1\%) lower than OPG, with an absolute difference of just 1.8 (SD=0.36), indicating that participants generally aligned their behavior to maximize AAG towards OPG. Interviews revealed strategies reflecting the OPG optimization process. For example, one participant noted, \textit{``Assign a profit to each task and try to maximize the total profit they could achieve given ADS's accuracy'' (P11)}, aligning closely with OPG calculations. Another participant mentioned considering strategies to \textit{``deliberately maximize their AAG throughout the collaboration process'' (P3)}. Additionally, with sufficient decision-making time, some participants followed strategies to maximize AAG by calculating decision probabilities based on ADS suggestions.

In addition to the correlation between AAG and OPG, we found that both metrics increased as ADS accuracy improved, suggesting that participants relied more on ADS suggestions to achieve higher AAG during collaboration. A Friedman non-parametric test revealed a significant effect of ADS accuracy on AAG across different tasks (Avoid Collision: $p < .001$; Overtake: $p < .001$; Route Selection: $p < .001$). In the next study, where AAG deviates from OPG, we will explore potential explanations from the drivers' decision-making perspective.

To rigorously assess the alignment between AAG and ground truth accuracy, we conducted a correlation analysis, where the ground truth was predefined based on ADS accuracy, though it was not disclosed to participants. As shown in Figure~\ref{fig:study2_correlation}, AAG demonstrated a strong correlation with the ground truth, with Pearson and Spearman coefficients of 0.77 and 0.78, respectively. Task-specific correlations also yielded high values: for collision avoidance, 0.72 (Pearson) and 0.59 (Spearman); for overtaking, 0.69 and 0.61; and for route selection, 0.96 for both coefficients. Accuracy-specific correlations further confirmed AAG's robustness: at 60\%, Pearson and Spearman coefficients were 0.69 and 0.68, respectively; at 90\%, 0.73 and 0.71; and at 99\%, 0.95 and 0.96. These results demonstrate AAG's effectiveness in aligning with ground truth while offering nuanced improvements over traditional metrics in capturing real-world detection accuracy.

\subsubsection{Effect of Task and ADS Accuracy on Decision-Making Patterns}

We analyzed participants' decision-making behavior using a similar approach to Section~\ref{sec:study1_cognition}. The following metrics were defined to quantify and explain these aspects.

\textbf{Follow rate}: defined as the number of times participants conformed to ADS suggestions divided by the total number of take-overs.

\textbf{Conservative rate}: defined as the number of times participants chose the safer option such as avoiding, not overtaking divided by the total number of take-overs. 

``Avoid Collision'' and ``Route Selection'' achieved the lowest and highest \textit{follow rate} respectively, which was associated the task's risk levels. However, because of the influence of ADS's accuracy, no significant effect of task was found on \textit{follow rate} ($\chi_2^2 = 2.34$, $p = .31$). The \textit{follow rate} increased significantly with higher ADS accuracy, showing a notable effect of ADS accuracy on the \textit{follow rate} ($\chi_2^2 = 61.5$, $p < .001$), with post-hoc Tukey tests indicating significance between the ``60\%'' and ``90\%'' accuracy levels ($p < .001$).

\begin{figure}[!htbp]
    \subfloat[Follow rate with different task and ADS accuracy.]{
        \includegraphics[width=0.47\textwidth]{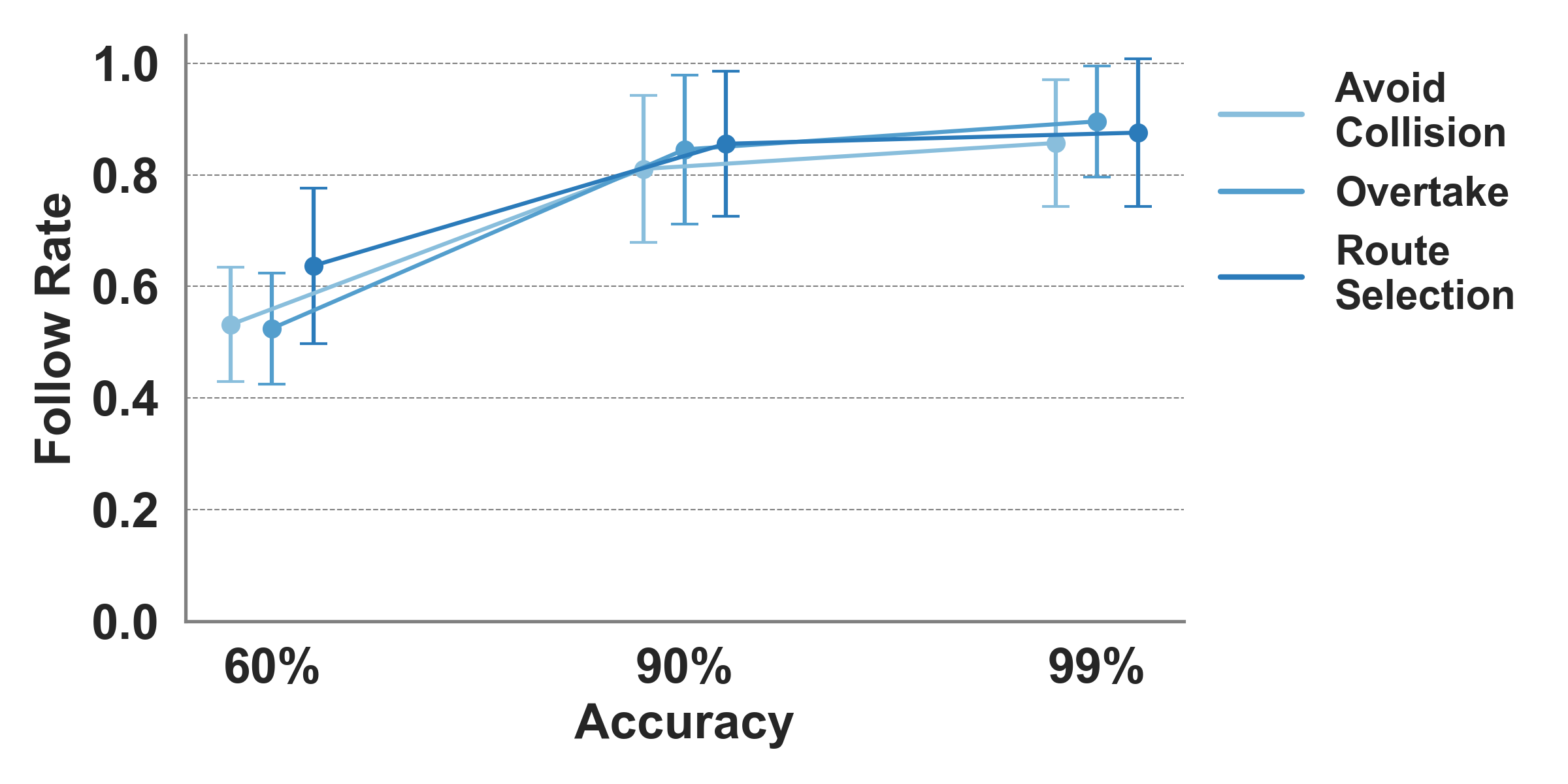}
    }
    
    \subfloat[Conservative rate with different task and ADS accuracy.]{
        \includegraphics[width=0.47\textwidth]{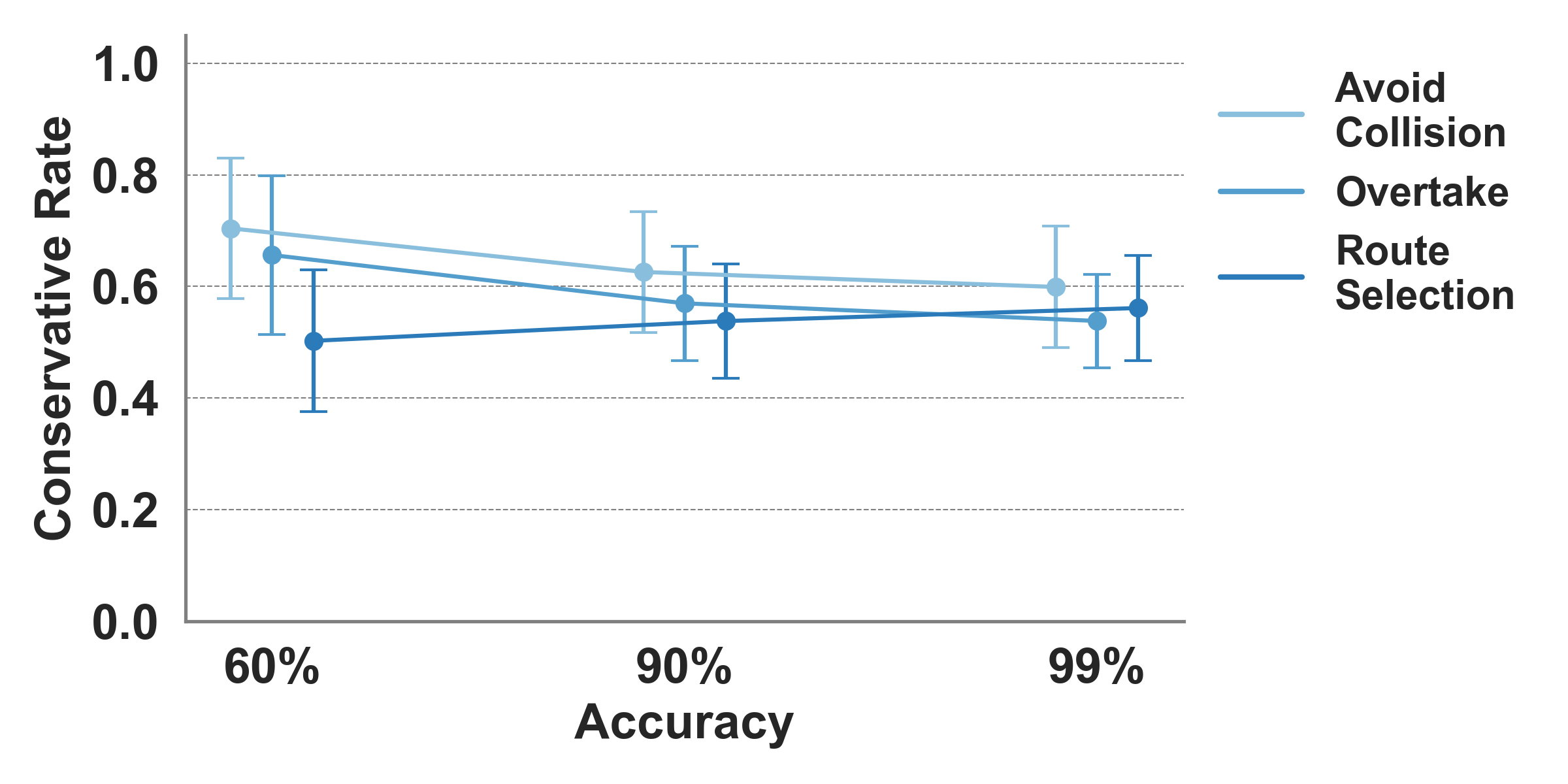}
    }
    \caption{Follow rate and conservative rate for different tasks and ADS accuracy. Error bar indicated one standard deviation.}
    \Description{This figure contains two graphs showing the follow rate and conservative rate under different tasks and ADS accuracy. Error bar indicated one standard deviation.The graphs uses ADS's accuracy as the horizontal axis and the follow rate (the first graph) or conservative rate (the second graph) as the vertical axis and use lines with different colors to stand for different tasks.}
    \label{study2-rate}
\end{figure}

The \textit{conservative rate} was significantly lower for ``Route Selection'', indicating that participants did not strongly favor conservative choices ($\chi^2_2 = 43.8$, $p < .001$, with post-hoc comparisons showing for ``Route Selection'' vs. ``Avoid Collision'', $p < .01$ and ``Overtake'', $p < .01$). In contrast, the effect of ADS accuracy on the \textit{conservative rate} was minimal, as its impact varied across tasks (as shown in Figure~\ref{study2-rate}). We also noticed a significant overall effect of ADS accuracy on the \textit{conservative rate}, which with higher ADS accuracy, participants tended to adhere to ADS's suggestions more, thus reducing the cases of choosing conservative choices ($\chi_2^2 = 6.70$, $p < .05$).

As shown in Figure~\ref{study2-rate}, for tasks with lower risk, participants exhibited a higher tendency to follow ADS suggestions, with the \textit{follow rate} significantly increasing as ADS accuracy improved, although it consistently remained below the reported accuracy. At the same time, the \textit{conservative rate} for ``Avoid Collision'' decreased, indicating a reduced tendency to choose avoidance when ADS advised otherwise, reflecting an increased reliance on ADS guidance. The \textit{conservative rate} did not significantly change with ADS accuracy across tasks, particularly in ``Route Selection'', indicating that participants did not display a strong risk aversion tendency regardless of ADS accuracy. Friedman test found that ADS accuracy had a significant impact on \textit{follow rate} across all tasks (Avoid Collision: $\chi^2_2=35.2$, $p < .001$; Overtake: $\chi^2_2=44.8$, $p < .001$; Route Selection: $\chi^2_2=14.4$, $p < .001$). Friedman test showed a significant effect of ADS accuracy on the \textbf{conservative rate} of ``Avoid Collision'' task ($\chi^2_2=6.81$, $p < .05$), but no significant effect was observed for ``Overtake'' ($\chi^2_2=5.69$, $p = .06$) or ``Route Selection'' ($\chi^2_2=2.35$, $p = .31$). The Nemenyi post-hoc test revealed a significant difference in \textit{follow rate} between 60\% and 90\% accuracy across tasks (post-hoc $p < .05$). No post-hoc significant differences of \textbf{conservative rate} was found for all tasks.  These findings confirm that participants were more likely to avoid risky choices when ADS accuracy was low and more inclined to follow ADS suggestions as accuracy increased.


\section{Study 3: Examining the Usage of AAG \& OPG with Different Decision Time}\label{sec:insufficient_time}

We evaluated the effectiveness of AAG and OPG under time constraints since most real-life decisions are made urgently~\cite{feldhutter2019effect}. 

\subsection{Participants and Apparatus}

We recruited 54 Chinese participants (31 males, 23 females) with an age ranging from 18 to 54 (M=28.1, SD=9.0) and an average driving experience of 5.7 years (SD=5.4) through snowball sampling~\cite{goodman1961snowball}. Participants were from varied occupations and with varied educational backgrounds. The apparatus was the same as in Study 2 and each participant received 100 RMB for their participation.

\subsection{Study Design}

We adopted a two-factor within-subjects design with \textbf{time} and \textbf{task} as factors. We chose the same tasks as in Study 1 and Study 2. Given that participants typically collaborate with high-accuracy ADS in reality, the ADS accuracy was set at 90\% following previous literature~\cite{azevedo2020real}. The accuracy also prevented participants to blindly follow without thinking. Decision time was defined as the duration from ADS suggestion to key pressing by the driver, which was potentially shorter than take-over time (TOT) that included initial motion control.

Previous research indicated varied take-over time~\cite{zhang2019determinants}. According to a survey by Zhang et al.~\cite{zhang2019determinants} and studies by Gold et al. and Zeeb et al.~\cite{gold2013take,zeeb2015determines}, drivers responded adequately with a take-over time of 5 seconds. Zhang et al. reported an average take-over time of 2.5-3 seconds~\cite{zhang2019determinants}. Based on this, we selected three time intervals: 0.5s, 1.5s, and 2.5s. In Study 2, we found drivers required no more than 3.5 seconds on average to complete their decision-making process~\cite{eriksson2017takeover}. Participants underwent 36 trials (3 decision times $\times$ 3 tasks $\times$ 2 suggestions $\times$ 2 repetitions), with each trial's procedure mirroring Study 2 but with restricted decision time. Each pair of trials per suggestion had one correct and one incorrect outcome to test decision accuracy.

During practice, a floating timer (600 $\times$ 300 pixels) was displayed in the top left corner, counting down from when ADS finished its suggestion. An alarm sounded when time expired.  No feedback was provided during the actual trials. Other settings, including ground truth calculation, decision time measurement and participant responses, were consistent with Study 2.

\subsection{Procedure}

Before the experiment, participants were introduced to the driving simulator, tasks, and the decision time constraints. Participants were given 3 minutes to familiarize themselves with the platform and practiced randomly generated trials until they met the required time threshold for 5 consecutive trials. The practice session last 9 trials on average. In the main experiment, participants completed 36 trials divided into two sessions of 18 trials each, similar to Study 2, with randomized trial order to prevent order effects. Each trial began with the system informing participants of the ADS accuracy and the specific decision time allowed. The ADS then drove to a set position and indicated the need for take-over. Participants made decisions within the allotted time upon the take-over. As in Study 2, the system proceeded to the next trial only after a decision was made. A 30-second break was given between sessions to minimize fatigue.

\subsection{Results}


We first examined whether participants adhered to the required time limits. Linear regression analysis showed a strong correlation between demanded decision time and actual decision time ($y = -0.05 + 1.04x$, $R^2 = 0.89$). Specifically, with a demanded decision time of 0.5s, the average decision time was 0.45s (SD=0.28s), indicating participants could decide within the required time limits. Because the study incorporated decision time and task as two factors, we adopted Aligned Rank Transform~\cite{wobbrock2011aligned} before performing the statistical test. We analyzed the users' decision behavior, reflected through AAG and OPG, and examined participants' irrational decisions.

\subsubsection{AAG Deviate from OPG with Insufficient Time}\label{sec:study3_OPG}

\begin{figure} 
    \subfloat[]{
        \includegraphics[width=0.47\textwidth]{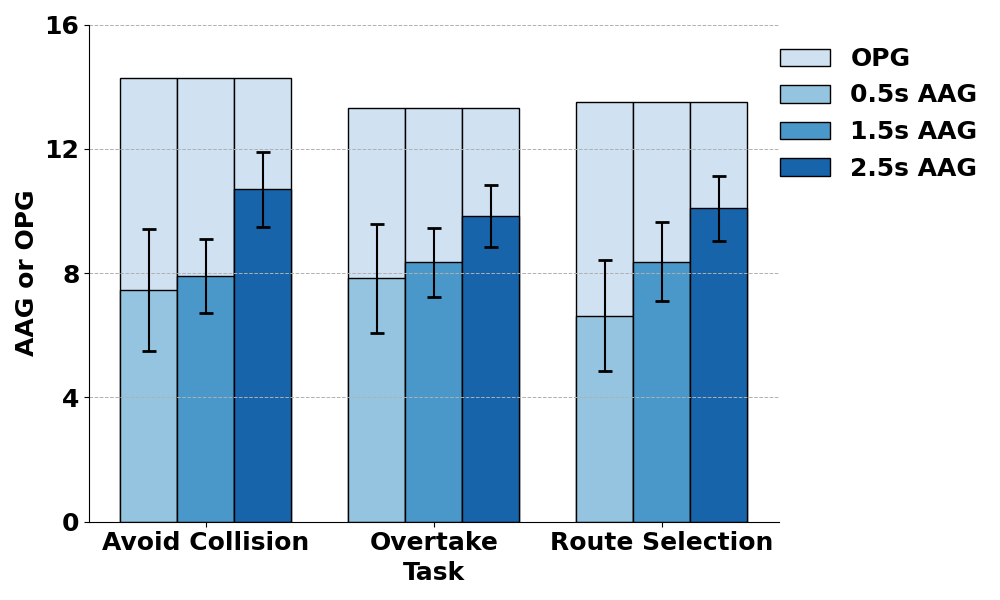}
        \label{fig:study3_opg}
    }
    
    \subfloat[]{
        \includegraphics[width=0.47\textwidth]{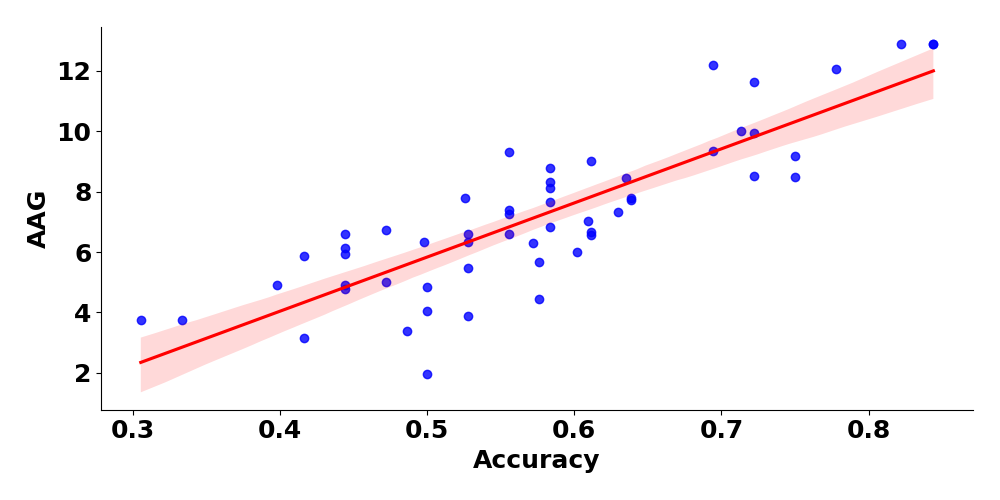}
        \label{fig:study3_regression}
    }
    
    \caption{(a) AAG and OPG with different tasks with different time. Errorbar indicated one standard deviation. (b) The regression plot of AAG and real world decision accuracy. The red shade indicated 95\% confidence interval.}
    \label{fig:study3_opg_aag}
\end{figure}

Figure~\ref{fig:study3_opg} displayed the AAG \& OPG across different tasks with varying decision time, revealing that drivers exhibited more irrational behavior as decision time decreased. Specifically, with a decision time of 0.5s, the gap between AAG and OPG was 48.8\% (SD=3.5\%), at 1.5s it was 38.4\% (SD=6.4\%), and at 2.5s, the gap reduced to 24.4\% (SD=1.7\%). A Friedman non-parametric test revealed a significant effect of decision time on AAG for ``Avoid Collision'', ``Overtake'', and ``Route Selection'' (all p < .001). This showed participants fall short of the optimal strategy when time is limited. Further analysis of the strategies drivers adopted is discussed in Section~\ref{sec:driver_strategy}. Participant feedback highlighted their strategies under time constraints, with one noting, \textit{``I have no time to think but to press what I hear. I think this would be a simple but effective strategy. I adopted this throughout the study.'' (P5)}

In addition to analyzing the AAG-OPG gap, we further evaluated the alignment of AAG with real-world decision accuracy (see Figure~\ref{fig:study3_regression}). Across all tasks, AAG demonstrated strong correlations with ground truth accuracy, with a Pearson coefficient of 0.85 and a Spearman coefficient of 0.84. Task-specific analyses revealed similarly high correlations: for avoiding collisions, Pearson and Spearman coefficients were 0.93 and 0.92, respectively; for overtaking, 0.89 and 0.88; and for route selection, 0.93 and 0.92. Results across different decision times were also consistent, with coefficients ranging from 0.87 to 0.88. These findings highlight AAG's alignment with traditional metrics, enhancing its usefulness in real-world tasks.

\subsubsection{Examining Drivers' Irrational Decisions through Behavioral Patterns}\label{sec:driver_strategy}

\begin{figure}[!htbp]
    \subfloat[Conservative rate.]{
        \includegraphics[width=0.46\textwidth]{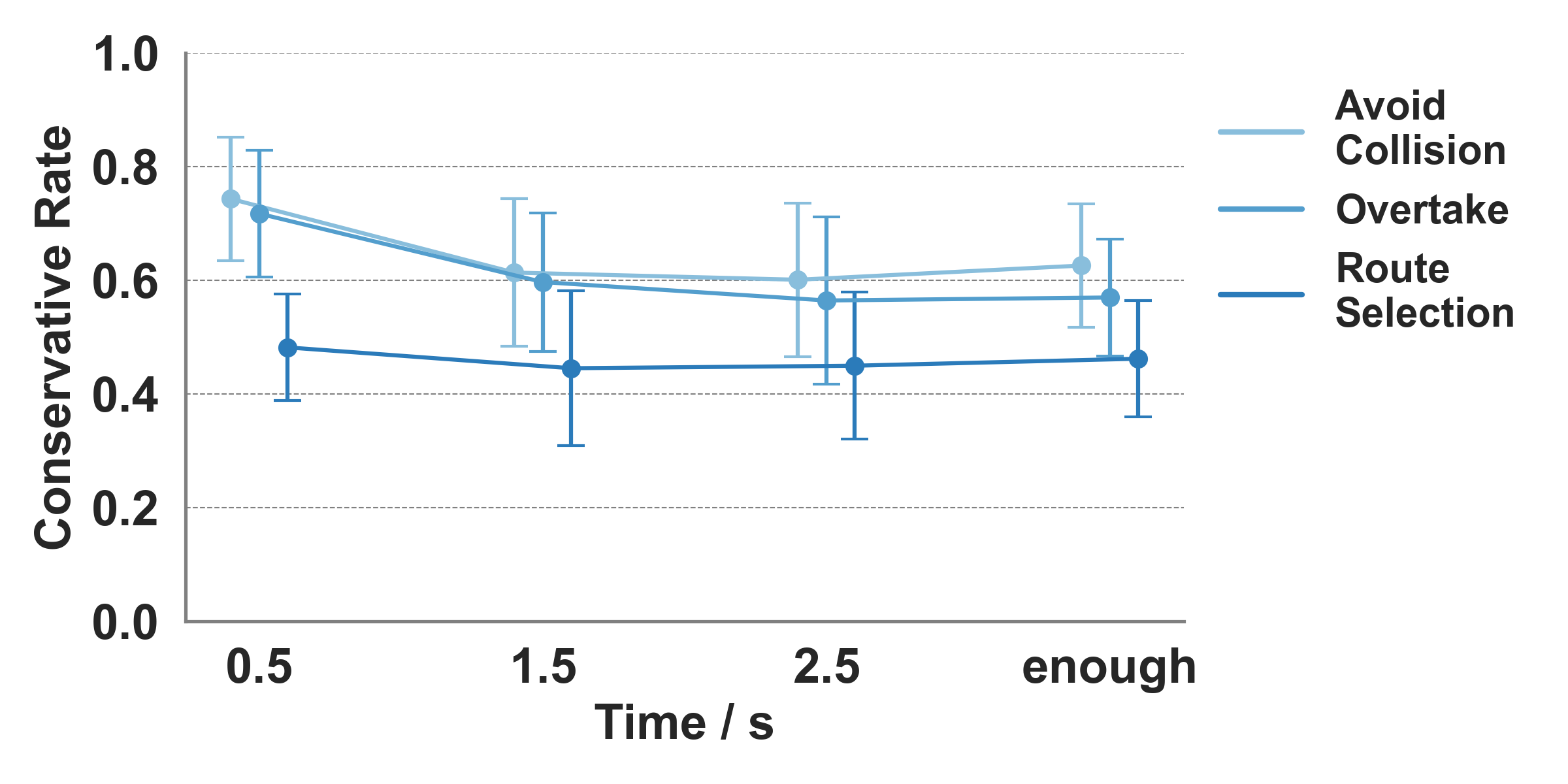}
    }
    
    \subfloat[Follow rate.]{
        \includegraphics[width=0.46\textwidth]{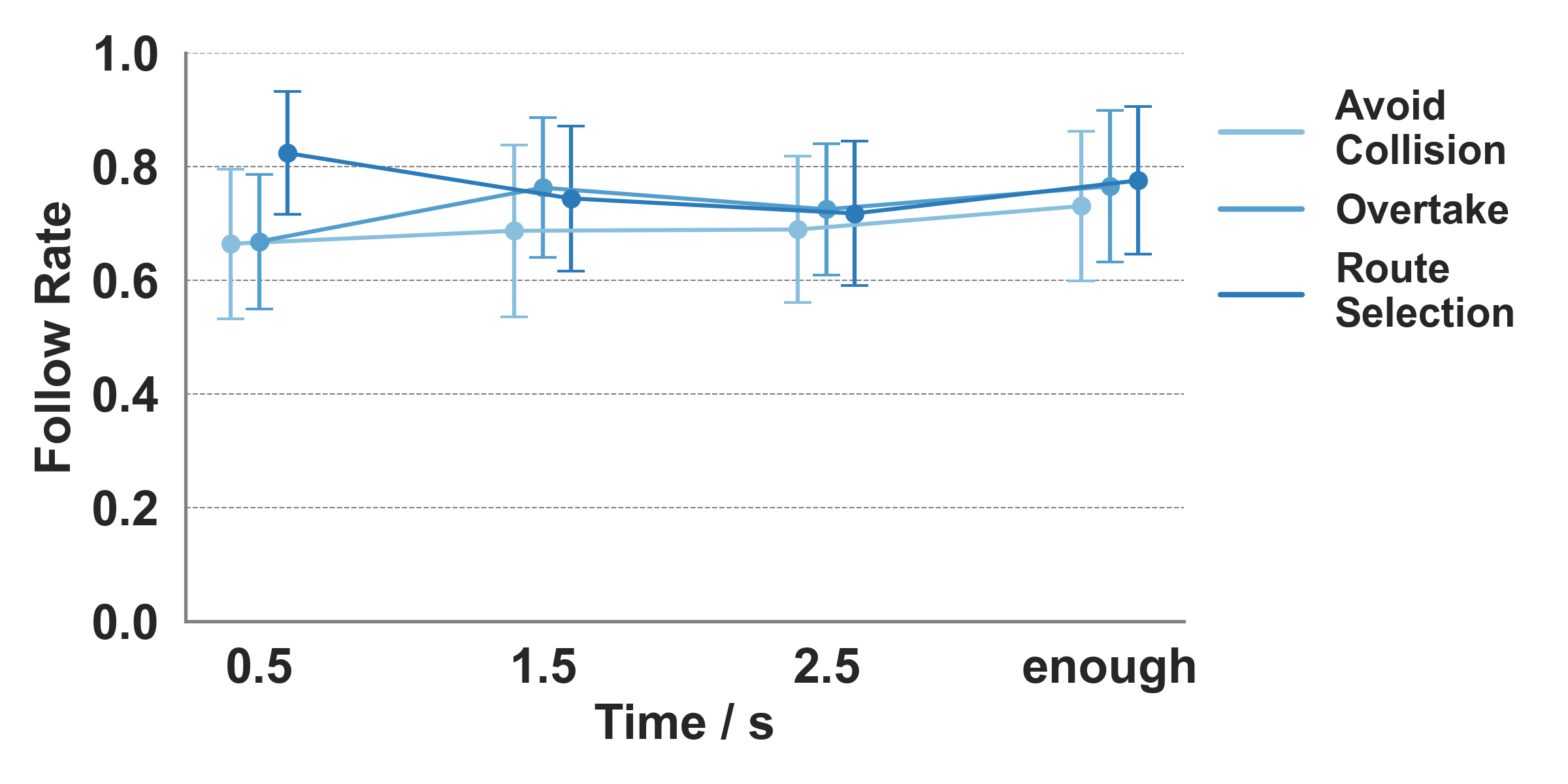}
    }
    \caption{\textit{Conservative rate} and \textit{follow rate} with different decision time for different tasks. Enough denoted the value in Study 2 with ADS accuracy 90\%. Errorbar indicate one standard deviation.}
    \label{fig:study3_comb}
\end{figure}

We attributed irrational decision-making to two key factors: irrational risk aversion~\cite{holt2002risk} and following. Figure~\ref{fig:study3_comb} shows the \textit{follow rate} and \textit{conservative rate} across tasks with varying decision times, and compares these results to those form Study 2 at a 90\% ADS accuracy. As decisoin time decreased, irrational patterns emerged across tasks. Specifically, the follow rate increased for the avoid collision task, while the conservative rate increased for the overtake and route selection tasks. Statistical analysis confirmed a significant effect of decision time on both metrics ($p < .01$ for all tasks). More specifically, the follow rate for avoid collision increased significantly as decision time decreased (post-hoc $p < .05$ for 0.5s vs. 1.5s and 2.5s), whereas conservative rates increased in the overtake and route selection tasks under time constraints (post-hoc $p < .05$ for 0.5s vs. 1.5s and 2.5s). These findings suggest that under time constraints, participants may either follow ADS recommendations more closely or adopt overly cautious strategies. However, both excessive conservation and blindly following potentially erroneous ADS suggestions can result in suboptimal decision-making, potentially leading to errors.

\section{Study 4: Verifying the Usage of AAG through an Intervention Study}

In previous studies, we calculated AAG and OPG to assess their effectiveness in reflecting decision-making quality and patterns. In this section, we aim to use interventions to improve drivers' decisions, as they often make irrational choices under time constraints.

\subsection{Design of the Intervention}\label{sec:design_principle}

Figure~\ref{fig:study3_comb} indicated with insufficient time, participants’ AAG often deviates from the OPG due to a lack of deliberation and a tendency towards risk aversion or following behavior~\cite{holt2002risk}. For example, participants frequently chose to avoid a collision, even when the suggestion was not to avoid it, especially in cases where the ADS suggestions were imperfect, leading to unnecessary time wasting. Insufficient time exacerbated this issue. To address this, our intervention is designed for scenarios where participants 1) have insufficient time to make decisions, and 2) face significant perceived gains and losses.

We considered several intervention forms: 1) providing different levels of information~\cite{zhou2021does,boelhouwer2019should,korber2018have,liu2018impact}, as information during interventions can be binary~\cite{boelhouwer2019should,korber2018have,liu2018impact} or multi-level~\cite{zhou2021does}; 2) allocating varying take-over times~\cite{braunagel2017ready}, given that in emergency scenarios, drivers often rely on ``System 1'' thinking, which can lead to irrational decisions~\cite{evans2003two}; and 3) providing take-over requests through different modalities, which have been shown to reduce drivers' reaction times~\cite{walch2015autonomous}.

To validate AAG's effectiveness and avoid other effects, we adopted the simplest form of intervention (form 1), focusing solely on varying levels of information. This approach minimized the influence of the intervention design itself and eliminated potential confounding factors related to modality and timing, allowing us to assess the impact of AAG as an additional information source.

\subsection{Participants and Apparatus}
We recruited 8 Chinese participants (4 males, 4 females) with a mean age of 23.6 (SD=1.3) from the campus. All were familiar with driving in the real car and have driving licenses. They have an average driving experience of 2.6 years (SD=1.2) and were all unfamiliar with driving in the simulator. We adopted the same experiment device as in the former studies. Each participant was compensated 150 RMB according to the local wage standard.

\subsection{Experiment Design}\label{sec:study4_experiment_design}

We adopted a one-factor within-subjects design with the \textbf{remind method} as the factor. We designed three remind methods: \emph{alert based on AAG (denoted as AAG)}, \emph{always alert (denoted as ``Base'')} and \emph{no alert (denoted as ``Null'')}. We adopted a simple alarm before ADS suggestions to minimize the extra influence of the reminding. The alarm consisted of three beeps with each length 0.2s. The interval between the beeps is 0.2s and the frequency of the beep was 2500Hz. For the AAG-based alert, the frequency and length of the beep remained unchanged, but the alarm was only triggered in cases where participants' AAG deviated from OPG. Four cases required intervention in the AAG alert group, forming a $2\times2$ matrix: 1) insufficient time (0.5s and 1.5s); 2) overtaking and route selection tasks where participants tended to blindly follow the ADS. The always alert group alerted participants with the beeps in each case. The no-alert group remained silent during the corresponding time.

Each remind method factor group involved 12 repeated trials per participant. To test more scenarios and prevent memorization effects, tasks, suggestions, ADS accuracy, and decision times were not fixed. Instead, trials were randomly generated with different factor levels. For each take-over task with corresponding ADS accuracy, the outcome was determined first. Then, the ADS report was decided based on the outcome and ADS accuracy. For example, with 90\% ADS accuracy, the ground truth matched ADS accuracy 90\% of the time. Measurements included:

\begin{itemize}
    \item \textbf{Correct Ratio (CR)}: Defined as the number of correct predictions divided by the total trials. ``Correct'' means if the ground truth allows overtaking, the participant overtakes.
    \item \textbf{AAG and OPG}: The ratio of AAG to OPG was also used to assess performance and rationality, aligning with the comparison AAG and OPG in the previous studies.
\end{itemize}

\subsection{Procedure}

Participants were given three minutes to familiarize themselves with the autonomous driving system. The study has three groups with randomized order. The tasks of each group was randomized, differed in take-over time (0.5s, 1.5s, 2.5s), the task (``Avoid Collision'', ``Overtake'', ``Route Selection'') and the ADS suggestions. Additionally, the environment was also randomized as in previous studies. For each group participants were asked to complete 12 randomized trials with each trial mirroring Study 3, except that the alarm was replaced. After the experiment, we conducted short interviews with each participant regarding their decisions, potential intervention cases, and the intervention's effect.

\subsection{Results}\label{7.3}

We prove the intervention through AAG is effective through analyzing quantitative metrics and providing subjective users' comments.

\subsubsection{Improvement Leveraging AAG}

\begin{figure}[!htbp]
    \subfloat[The ratio of AAG and OPG.]{
        \includegraphics[width=0.42\columnwidth]{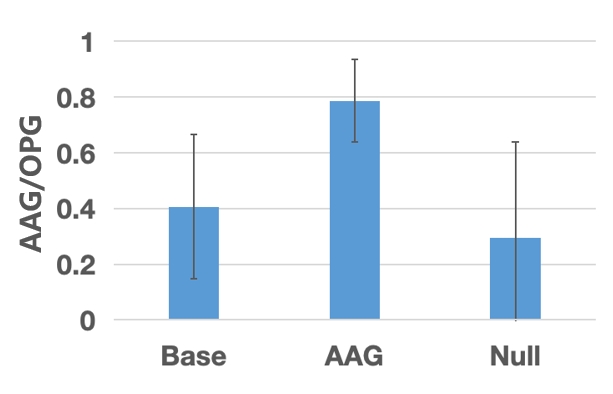}
        \label{fig:aag_ratio}
    }
    \subfloat[CR of participants.]{
        \includegraphics[width=0.47\columnwidth]{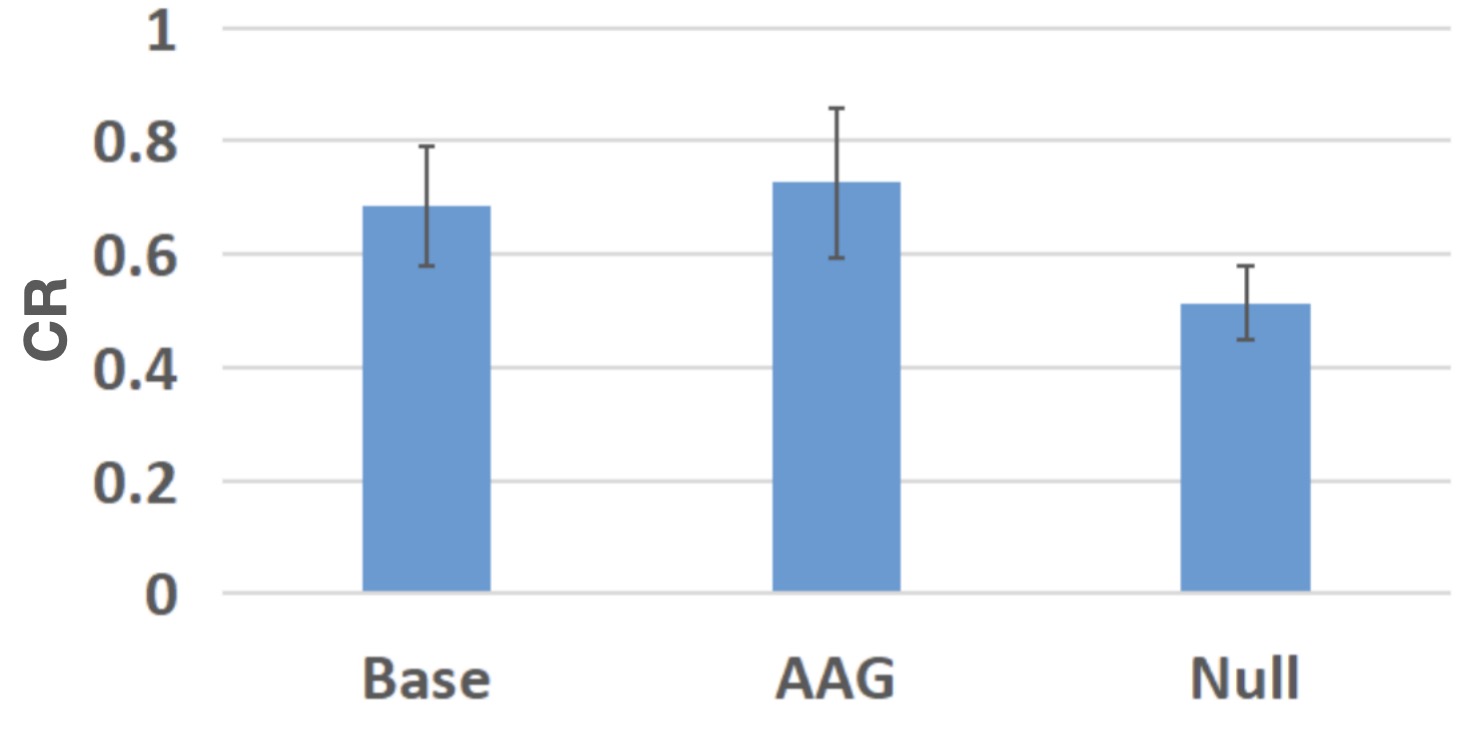}
        \label{fig:correct_ratio}
    }
    \caption{(a) The ratio of AAG and OPG, (b) the CR of participants for different intervention settings. The error bar showed one standard deviation.}
    \label{fig:aag_OPG}
\end{figure}

We examined the AAG participants achieved. Figure~\ref{fig:aag_ratio} plotted the ratio of AAG and OPG, whose difference reflected the cases where participants do not follow the vehicles. Figure~\ref{fig:correct_ratio} plotted the participants' correctness likelihood for different intervention settings, which reflected both the final decision accuracy and the correctness ratio of participants. The AAG approached OPG (79.0\%) with the intervention design compared with ``Base'' (40.5\%, +95.1\%) and Null (29.0\%, +172.4\%), and the correctness likelihood also increased correspondingly (+4.3\% compared with ``Base'' and +38.4\% compared with ``Null''). The results indicated that with AAG \& OPG, ADS could better predict participants' behavior, take intervention and participants would act better correspondingly. This echoed Walch et al.'s \cite{walch2017car} advocate of mutual prediction.

\subsubsection{Subjective Feedback}

We interviewed participants about whether the intervention was effective. Participants generally agreed that presenting beeps would raise their attention, for which cases they would \textit{``decide with more thinking process''} and \textit{``react faster''} (P2). Besides, they also mentioned that continuous beeping could \textit{``affect their driving experience, which made them indifferent to the intervention''} (P4). Thus, they agreed that \textit{``the beeping should be presented in cases where they should focus more and when ADS perform worse, rather than in all cases''} (P1). This coincided with our design principle (see Section~\ref{sec:design_principle}) and the findings regarding AAG (see Figure~\ref{fig:study3_opg}). During the intervention, AAG is used to select these cases where driver performance was suboptimal. For those cases, participants agreed that additional focus and cognitive effort were required, supporting the need for intervention. To avoid overwhelming the participants, they echoed the number of such cases should remain manageable~\cite{zeeb2015determines}.

Furthermore, participants' expectations for when the intervention was most needed mirrored our design. Participants revealed that the cases that needed the intervention the most was 1) when the decision time is insufficient, 2) when the gain and loss was larger in value. These conditions facilitated the improvement of participants' CR and led to higher ratios of AAG \& OPG. We also find for the above cases, drivers' responses did tend to be irrational (see Figure~\ref{fig:study3_opg}). 

\section{Discussion}

\subsection{Significance of AAG \& OPG over Existing Works}

AAG \& OPG originated from perceived gain and loss, which in economics is measured using subjective scales~\cite{mcgraw2010comparing,gal2018loss}. Unlike previous studies on tasks like maze solving~\cite{vasconcelos2023explanations} or gambling~\cite{lin2023gain} tasks, we focus on autonomous driving, tailoring the framework to account for safety and time-related factors. While prior research used similar subjective measurements to assess risk perception in autonomous driving~\cite{brell2019scary,li2019no}, they did not link perception to decisions. AAG \& OPG improved decision accuracies by incorporating task-specific perceptions, such as safety and time considerations, aligning with risk perception literature~\cite{sourelli2021objective}. Within this framework, different behaviors are assigned varying levels of importance based on the severity of their potential consequences, aligning with the intuitive understanding that avoiding collisions is far more critical than regular driving. Study 1 (see Table~\ref{study1_task}) and subsequent studies (see Figure~\ref{fig:study2_aag} and~\ref{fig:study3_opg}) show that AAG \& OPG aligns closely with real-world accuracy and further priortize safety. 

AAG \& OPG do not rely on consecutive bio-signals~\cite{yu2021takeover} or ground truth feedback~\cite{du2020predicting}, enhancing their real-world applicability. AAG \& OPG are potentially applicable to various similar take-over tasks (e.g., avoiding obstacles~\cite{navarro2016obstacle}, pedestrians~\cite{schratter2019pedestrian}, lane changing~\cite{smirnov2021game}), as well as different take-over types, including partial, assisted~\cite{gruden2023assisted}, and driver-initiated~\cite{10.1145/3003715.3005456} driving. These tasks, similar to our work, has potential safety risks~\cite{holt2002risk} and time saving effects~\cite{steck2018autonomous} which influenced participants' decisions. 

In Study 2 and 3, we examined how constructing factors like ADS accuracy, decisions, tasks and time influence AAG (see Figure~\ref{fig:study2_aag} and~\ref{fig:study3_opg}), while also modeling human decision-making. Regarding time, Gold et al.~\cite{gold2013take} found that limited time negatively affects driver take-overs, though their analysis focused primarily on control. Our findings extend this by showing that insufficient time also leads to imperfect human decisions, suggesting that ADS could proactively offer alternative suggestions, as discussed in Section~\ref{7.3}. The past work also highlighted the risk aversion~\cite{holt2002risk} behavior of the human's perception towards ADS~\cite{wang2019risk}. Building on this, we find risk aversion towards road conditions leads to more conservative decision-making behavior (See Figure~\ref{fig:study3_opg} and~\ref{fig:study3_comb}). If ADS can recognize these behavior patterns, it could enhance collaborative
 decision-making between humans and ADS (See Section~\ref{sec:discussion_design}).

\subsection{Generalizability of the Results}

To model drivers decision quality in critical situations like collisions, our study employed driving simulators to comply with ethical and safety considerations~\cite{mcgehee2010perception}. Simulators are widely validated tools for evaluating driver behavior and risk perception, facilitating safer and controlled repeatable experiments that maintain generalizability across prior research~\cite{carsten2011driving, wang2022studying, zhang2019determinants}. In real-world scenarios, haptic cues and contextual factors can increase driver anxiety and cognitive load, impairing rational decision-making and widening the deviation between AAG and OPG, necessitating
decision quality improvement. Adjusting alert strategies based on AAG \& OPG deviations can better guide drivers toward rational decisions, demonstrating the practical value of AAG \& OPG in real-world driving.

To enhance the robustness of the results, we varied road conditions, including urban, rural, and highway environments, as well as traffic density and visibility. Real-world driving is influenced by factors like weather, time of day and road types, all of which impact risk perception and decision-making. For instance, adverse weather or nighttime driving can heighten caution and change risk responses~\cite{repa1982effect}. Drivers differences, such as risk aversion, also impact how they respond to warnings~\cite{ulleberg2003personality}. By including these variations, out findings are more broadly applicable across different driving contexts.

Our studies primarily focused on the development and validation of the AAG \& OPG metrics, while acknowledging that potential applications of AAG \& OPG could leverage multi-modal alerts (e.g., visual or haptic feedback) to enhance decision-making~\cite{faltaous2018design,telpaz2015haptic}. Voice alerts, commonly used in take-over scenarios~\cite{wang2022speech, bazilinskyy2018take, petermeijer2017take}, served as the primary alert modality in our study. Different alert types affect cognitive load and decision-making~\cite{merenda2017did,wen2024adaptivevoice}: auditory alerts impose lower cognitive demands whereas multimodal alerts were informative but potentially demanding. Balancing these alert forms based on AAG-OPG deviations, cognitive load, and situational awareness could improve driver responses~\cite{wen2024adaptivevoice, zeeb2016take, bouyer2017inducing}.

\noindent \textbf{Multimodal Alert Experiment} We conducted a preliminary study to explore the potential applications of AAG \& OPG with multimodal alerts. The intervention employed a pop-up display, consistent with prior studies~\cite{yun2020multimodal,li2024impact,yun2018experimental}. Since most alerts adopt visual and auditory forms~\cite{deng2024design}, we focused on these modalities and their combinations. The pop-up display refreshed at 3Hz~\cite{li2024impact,yun2020multimodal}, in line with previous work. We primarily used suggestions with icons as the interface content similar to Walch and Deng's work~\cite{walch2015autonomous,deng2024design}, aligning with common guidelines~\cite{yun2020multimodal}. We designed four experimental conditions: (1) A multi-modal display using both visual and auditory channels for urgent cases, switching to auditory-only for less urgent cases to reduce cognitive load. (2) A constant multi-modal display for all cases where AAG deviates from OPG, similar to Study 4. (3) A single-channel visual display. (4) No additional display. Four participants (1 male, 3 females; mean age=32.3, SD=9.8; mean driving experience=11.3 years, SD=8.3) completed four counter-balanced sessions, with 12 trials per condition. We adopted the platform similar to previous studies and analyzed decision accuracy as well as AAG-OPG alignment. 

Results showed that the multi-modal display (group 1) achieved a correctness likelihood of 8.5\% higher than the auditory-only group (group 2) and 46.2\% higher than the no-display group (group 4). Multi-modal alerts also improved AAG-OPG alignment (group 1: 79.4\%, group 2: 70.1\%, group 3: 49.3\%, group 4: 35.0\%), reflecting improved decision quality. These findings suggest that AAG can align with OPG to improve driver decisions. While our study used auditory alerts, the AAG modeling is independent of alert type, making these results applicable to other alert systems.

\subsection{Design Implications}\label{sec:discussion_design}

AAG \& OPG can assess drivers' decision-making quality during take-overs, with key implications as follows: 

\noindent \textbf{Modeling decision quality in human-AI collaboration.} 
The metrics and frameworks developed in Section 3 and Study 1 for evaluating decision quality in take-overs can be applied to broader human-AI collaboration tasks, improving mutual understanding between human and AIs~\cite{ma2023should}. In healthcare decision support, framing treatment options as gains (e.g., improved outcomes) or losses (e.g., risks) can assist physicians in balancing risks and benefits. Similarly, in financial decision-making, such as with robo-advisors, framing investment choices as potential gains (e.g., high returns) or losses (e.g., financial risks) can help users align decisions with their risk tolerance. Leveraging AAG \& OPG, the AIs could better estimate human's decision quality, facilitating effective collaboration in complex decision environments.

\noindent \textbf{ADS could reduce take-over requests by predicting compliance of drivers.} 
As shown in Study 2 and 3, AAG enables ADS to predict whether drivers will comply with its instructions. When high compliance is expected, ADS might autonomously make decisions, reducing drivers' intervention and mental load~\cite{park2020driver}. Since drivers may have different preferences, the system could gradually adapt the level of automation based on the driver's recent decision patterns and dynamic AAG feedback, ensuring the system aligns with individual preferences. This approach is critical under time pressure, where drivers may over-trust the ADS (see Figure~\ref{fig:study3_comb}).

\noindent \textbf{Customizing Alert Timing and Content.} 
Study 3 demonstrates that ADS can adjust alerts according to task context and historical AAG-OPG deviation. When deviation is high, ADS should give more attractive alerts to gather drivers' attention. For high-risk tasks, drivers tend to make conservative decisions. Accuracy reports should highlight the system's risk assessment capabilities to counteract driver risk aversion.

\noindent \textbf{Integrating AAG and OPG to Tailor Multimodal Take-over Alerts.} 
Study 4 shows that alerts could adapt dynamically to the quantified deviations between AAG and OPG, tailoring their modality, intensity and informational content to the drivers' current decision-making state~\cite{walch2015autonomous,boelhouwer2019should}. When AAG indicates irrational decision tendencies under time constraints (e.g., over-reliance on ADS recommendations)~\cite{roche2018should}, multimodal feedback could be added to deliver concise yet salient prompts~\cite{yun2020multimodal}, directing attention without overwhelming cognitive load. Conversely, when drivers' AAG aligns closely with OPG, low intensity visual signals could suffice, reducing unnecessary distraction.


\section{Limitation and Future Work}

We acknowledge that simulated driving scenarios in our study lack some of the authenticity in real-world driving, potentially influencing participants' decisions. The absence of real-world sensory feedback, such as tactile acceleration cues~\cite{geitner2019comparison}, can decrease drivers' attention and consume less cognitive load, potentially leading to better performance in the simulator compared to actual driving. Haptic cues in real-world driving may heighten engagement, increase cognitive load and affect response times~\cite{wierwille1983driver,repa1982effect}. Without motion feedback, simulators may alter users' speed perception and reduce urgency, which may lead to smaller discrepancies between AAG and OPG during critical moments due to less irrational conservative decision-making. This further highlighted the problems that under real conditions, the deviation of AAG and OPG is larger. We propose that future work could conduct on-road experiments with safety measures in place. 

In our study, we sought to replicate real-world cognitive demands by incorporating dynamic stimuli such as traffic signals, vehicle interactions, and varying road conditions. These features aimed to replicate the cognitive demands of real-world driving and provide a more accurate representation of decision-making under complex conditions. However, it is important to note that simulators typically simplify the range of stimuli compared to real-world environments, which may result in reduced cognitive load and underutilization of cognitive resources. 

Our participants in Study 2 and 3 were recruited from a single region due to the in-lab study setting. While traffic laws did not significantly affect decision-making as the studies focused on cognitive processes and value judgments that are not tied to specific regulations, regional driving habits may introduce some variability. Future research should test our metrics across different cultures and traffic laws for broader applicability.

Furthermore, while other factors such as non-driving related tasks may influence collaborative decision-making, we focused on examining the most critical elements, treating others as random variables to strengthen the robustness of the results. Further research can explore additional factors that may affect the decision process.

\section{Conclusion}

In semi-automated vehicles, collaboration between humans and ADS is critical for passenger safety, particularly during the take-over process. We focus on the decision-making process where ADS provides suggestions and the driver makes the final decision. To quantitatively  model this process, we introduce AAG and OPG, calculated using a weighted combination of perceived gains and losses across different scenarios within a specific task. Our research began with a large-scale survey (N=315) using a Thurstone-scale questionnaire to measure perceived gains and losses as parameters for AAG and OPG. We then conducted two studies (N=54 each) to assess their utility under varying conditions, such as ADS accuracy and decision time.

Our studies confirm that AAG aligns with ground truth accuracy. With sufficient time, drivers tend to maximize AAG, tending towards OPG, with only 15.4\% deviation. Conversely, under time constraints, drivers adopted intuitive patterns, such as avoiding or simply following ADS suggestions, regardless of their variation. These patterns, reflecting risk aversion and following behaviour, could be mitigated through AAG-based interventions. Furthermore, we demonstrate AAG's application in intervention studies with audio alerts (N=8) and multimodal alerts (N=4). Preliminary findings suggest that both alert types improve the alignment of AAG and OPG, as well as correctness likelihood. We envisioned the application of AAG in other take-over tasks and forms, offering insights for future mutual understanding of ADS.

\begin{acks}
This work was supported by the Natural Science Foundation of China (NSFC) under Grant No. 62132010, 62472243, and 2022YFB3105201.
\end{acks}

\bibliographystyle{ACM-Reference-Format}
\bibliography{sample-base}

\appendix

\section{Questionnaire Items in Study 1}\label{sec:questionnaire}

Before the questionnaire, we added the explanation towards ``human's decision'', ``gain'' and ``loss'' as follows:

``The following question pertains to decision-making in various tasks within autonomous driving, involving two relatively new concepts: losses and gains. Losses refer to the disadvantages you believe would result from choosing a particular decision for various reasons. The greater the disadvantage, the larger the absolute value of the loss, with losses represented as negative values. Gains, on the other hand, represent the benefits you believe would result from choosing a specific decision for various reasons, with larger benefits corresponding to larger positive values. It is crucial to consider the relative values of gains and losses, specifically the relative relationships between gains and losses across different choices and scenarios. Please refrain from providing relative relationships that contradict normal logical reasoning.''

``Additionally, the risk in the first scenario is greater than that in the second scenario, and the risk in the third scenario is the smallest. You need to take this factor into consideration. Furthermore, the autonomous driving system's recognition accuracy for Level 3 autonomous vehicles is very high, and you can assume no difference in recognition accuracy across tasks. Lastly, please assess the absolute values of gains and losses reasonably, ensuring that their absolute values can be compared across tasks and actual choices. For gains/losses, you only need to provide a single value, with positive numbers representing gains (0 to 10) and negative numbers representing losses (-10 to 0). In the subsequent 12 questions, each box should correspond to a decimal or integer between -10 and 10.''

For each task, we added the corresponding explanation before the questionnaire item as follows:

``Assuming you are driving an autonomous vehicle at Level 3, these vehicles are characterized by not requiring active driving from you, but the vehicle may request you to take control of the driving at any time. Therefore, you still need to pay attention to the surrounding road conditions. In this scenario, the first task you encounter is avoiding oncoming traffic at a side intersection. Evading the oncoming vehicle can prevent a traffic accident, but if there is no actual risk of collision, not avoiding it could save time. You need to assess the gains or losses under this decision based on your response to the advice given by the autonomous driving system. The gains here could be in terms of time, fuel consumption, and energy saved by not avoiding, while losses could involve economic, property, and personal safety implications in the event of a collision. You may evaluate specific gains and losses based on the situation.''

``Continuing with the assumption that you are driving an autonomous vehicle at Level 3, in this scenario, the second task you face is deciding whether to overtake another vehicle. Overtaking may yield significant time gains, but there is also a very small probability of a traffic accident occurring during the maneuver. You need to assess the gains or losses under this decision based on your response to the advice given by the autonomous driving system. The gains here could be in terms of time saved by overtaking, while losses could involve fuel consumption, energy, and potential losses due to a traffic accident caused by overtaking. The likelihood and severity of a traffic accident in the overtaking scenario are much lower than in the first task. You may evaluate specific gains and losses based on the situation.''

``Continuing with the assumption that you are driving a Level 3 autonomous vehicle, in this scenario, the third task you face is choosing the forward route. Opting for a shorter route may offer the potential for higher speed, but it also entails the risk of encountering traffic congestion. Choosing a longer route, although covering more distance, might result in a shorter overall travel time. You need to assess the gains or losses under this decision based on your response to the advice given by the autonomous driving system. The gains here could be in terms of time saved by choosing the shorter route, while losses could involve fuel consumption, energy, and potential time losses if the shorter route encounters traffic congestion. You may evaluate specific gains and losses based on the situation.''

We provided the four ratings in thurstone entry after each description.

\section{The Procedure of Study 2 and Study 3}

In Study 2, the procedure of a single trial is as follows:
\begin{itemize}
    \item The system loads the relevant map and scene. The participant, seated in the driver’s seat, views the street through the windshield.
    \item The system reports its accuracy through voice. Then it reminded the participant that he or she has enough time to decide.
    \item The system switches to autonomous driving mode and accelerates from a stop.
    \item After a random interval of 15 to 60 seconds, the system notifies the participant of an urgent task and provides a suggestion via voice, requiring a decision.
    \item The image freezes at the moment the autonomous system requests a take-over. After the decision is made, the system transitions to the scene for the next experimental round.
\end{itemize}

In Study 3, during practice stage, the procedure of a single trial is as follows:
\begin{itemize}
    \item The system loads the relevant map and scene. The participant, seated in the driver’s seat, views the street through the windshield.
    \item The system announces its accuracy as 90\% via voice and informs the participant that they must make a decision within 0.5, 1.5, or 2.5 seconds.
    \item The system switches to autonomous driving mode and accelerates from a stop.
    \item After a random time between 15--60s, it informs the participant of the urgent task and then its suggestion through voice. It requires the participant to make the decision. Along with the suggestion, a pop-up window ($300\times600 pixel$ consisting a countdown of 0.5s/1.5s/2.5s would show on the top left of the window).
    \item The image freezes when the system requests a take-over. If no decision is made before the time expires, an alarm sounds to indicate a timeout. After the decision or timeout, the system transitions to the next round’s scene.
\end{itemize}

During test stage, the procedure of a single trial is as follows:
\begin{itemize}
    \item The system loads the relevant map and scene. The participant, seated in the driver’s seat, views the street through the windshield.
    \item The system announces its accuracy as 90\% via voice and informs the participant that they must make a decision within 0.5, 1.5, or 2.5 seconds.
    \item The system switches to autonomous driving mode and accelerates from a stop.
    \item After a random interval of 15 to 60 seconds, the system notifies the participant of an urgent task and provides a recommendation via voice, requiring a decision.
    \item The image freezes when the system requests a take-over. After the decision is made, the system transitions to the next round's scene.
\end{itemize}








\end{document}